\documentclass[journal]{IEEEtran}

\usepackage{amsmath,amsfonts,amssymb}
\usepackage{graphicx}
\usepackage{cite}
\usepackage{multirow}
\usepackage{algorithm}
\usepackage{algorithmic}
\usepackage{subfigure}
\usepackage{multirow}
\usepackage{url}
\usepackage{color}
\usepackage{array}
\usepackage{subfig}
\usepackage{tabularx}
\usepackage{balance}

\graphicspath{{./figsJournal/}}

\IEEEoverridecommandlockouts

\mathchardef\mhyphen="2D

\newcommand{\PRayleigh}{\mathrm{(P1\mhyphen R)}}
\newcommand{\PMISO}{\mathrm{(P1\mhyphen MISO)}}
\newcommand{\PMassive}{\mathrm{(P1\mhyphen large\mhyphen M)}}
\newcommand{\Hone}{\mathbf {H}_{\mathcal{N}_1}}
\newcommand{\Htwo}{\mathbf H_{\mathcal{N}_2}}

\newtheorem{lemma}{Lemma}

\newtheorem{proposition}{Proposition}

\newcommand{\Hbar}{\mathbf {\bar H}}
\newcommand{\Hw}{\mathbf H_{\mathrm{w}}}
\newcommand{\Hwone}{\mathbf H_{\mathrm{w},\mathcal{N}_1}}
\newcommand{\Hwonetd}{\tilde{\mathbf H}_{\mathrm{w},\mathcal{N}_1}}
\newcommand{\Hwonehat}{\hat{\mathbf H}_{\mathrm{w},\mathcal{N}_1}}
\newcommand{\Hwtwo}{\mathbf H_{\mathrm{w},\mathcal{N}_2}}

\newcommand{\xE}{\mathbb{E}}

\newcommand{\tr}{\mathrm{tr}}
\newcommand{\ttr}{\text{tr}}

\newcommand{\Qnet}{\bar Q_{\text{net}}}

\begin{document}
\title{Optimized Training Design for Wireless Energy Transfer}
\author{Yong~Zeng,~\IEEEmembership{Member,~IEEE,} and Rui~Zhang,~\IEEEmembership{Member,~IEEE}
\thanks{The authors are with the Department of Electrical and Computer Engineering, National University of Singapore.
(e-mail: \{elezeng, elezhang\}@nus.edu.sg).}
}

\maketitle

\begin{abstract}
Radio-frequency (RF) enabled wireless  energy transfer (WET), as a promising solution to provide cost-effective and reliable power supplies  for energy-constrained wireless networks, has drawn  growing  interests  recently. To overcome the significant propagation loss over distance, employing multi-antennas at the energy transmitter (ET) to more efficiently direct wireless energy to desired energy receivers (ERs), termed \emph{energy beamforming}, is an essential technique for enabling WET. However, the achievable gain of energy beamforming crucially depends on the available channel state information (CSI) at the ET, which needs to be acquired practically. In this paper, we study the design of an efficient channel acquisition method for a point-to-point multiple-input multiple-output (MIMO) WET system by exploiting the channel reciprocity, i.e., the ET estimates the CSI via dedicated reverse-link training from the ER. Considering the limited energy availability at the ER, the training strategy should be carefully designed so that the channel can be estimated with sufficient accuracy, and yet without consuming excessive energy at the ER. To this end, we propose  to maximize the  \emph{net}  harvested  energy at the ER, which is the average harvested energy  offset by that used for channel training. An optimization problem is formulated for the training design over  MIMO Rician fading channels, including the subset of ER antennas to be trained, as well as the training time and power allocated. Closed-form solutions are obtained for some special scenarios, based on which useful insights are drawn on when training should be employed to improve the net transferred energy in MIMO WET systems. 
\end{abstract}


\section{Introduction}
Wireless energy transfer (WET) has attracted significant interests recently due to its great potential to provide cost-effective and reliable power supplies for energy-constrained wireless networks, such as sensor networks \cite{502}. Generally speaking, WET can be implemented based on either the ``near-field'' techniques including inductive coupling and magnetic resonance coupling, or the ``far-field'' electromagnetic (EM) radiation (see e.g. \cite{525,534} and references therein), which are suitable for short-range (less than tens of centimeters), middle-range (say, a couple of  meters), and long-range (up to tens of kilometers) applications, respectively. In this paper, we focus on the EM radiation or radio-frequency (RF) enabled WET, where dedicated energy-bearing signals are transmitted from the energy transmitter (ET) to the energy receiver (ER) for energy harvesting. Besides longer transmission distance, RF-enabled WET enjoys other advantages over the two near-field techniques, such as more flexibility in deployment, smaller receiver size, easier implementation of energy multicasting  to a large number of wireless devices distributed in a wide area, and the more convenience to integrate with simultaneous information transmission \cite{478}.

The history of microwave enabled wireless energy transmission can be traced back to the early work of Heinrich Hertz in 1880's \cite{505}. Later in 1960's, RF-based WET in free space was extensively studied, with two well-known applications in microwave-powered helicopter \cite{506} and solar power satellite (SPS) \cite{507}. In these WET systems, high transmission power and large antenna apertures were used to overcome the propagation loss for large power reception. During the past decade, the interest in WET has reemerged, which was mainly driven by the increasing need to provide low-cost  and perpetual energy supplies to wireless devices with relatively low-power requirements, such as those in wireless sensor, body and personal networks, to replace the traditional battery-powered solution that requires manual battery replacement and/or recharging.  Besides, RF-based WET is being actively investigated in wireless communication systems such as cellular networks \cite{508}, cognitive radio networks \cite{510,509}, and relay networks \cite{535,511}.  The development in WET technology has also opened up several interesting new applications, e.g., simultaneous wireless information and power transfer (SWIPT) (see e.g. \cite{478,514,521,522}), where energy and information are jointly transmitted using the same RF waveform,  and wireless powered communication networks (WPCN) (see e.g. \cite{515,516,518}), where the communication devices are fully or primarily powered by means of WET.

One critical issue for the design of RF WET systems is to enhance the end-to-end energy transfer efficiency, which is the ratio of the energy harvested by the ER to that spent at the ET. Generally speaking, omni-directional energy transmission could result in very poor efficiency due to the fast power attenuation over distance. On the other hand, uni-directional transmission, which can be achieved by designing the antenna radiation pattern to form sharp beam towards the ER, could significantly enhance the energy transfer efficiency; but it generally relies on the existence of a clear line-of-sight (LOS) link between the ET and ER. For indoor WET applications in rich scattering environment, e.g., wireless sensors for smart buildings \cite{517}, the LOS link may not exist, or could be weak as compared to the scattered signal components. In such scenarios, rather than focusing all energy along the LOS link, it is preferable to balance the signal power and phases over the different paths so that the signals could add constructively at the ER for maximum energy delivery. To this end, multi-antenna based {\it energy beamforming} has been proposed \cite{478}, where the energy-bearing signals are weighted at the multiple transmit antennas before transmission to achieve energy beamforming gains. In this regard, massive multiple-input multiple-output (MIMO)  technique \cite{374,497} is particularly suitable for RF WET, where a very large number of antennas are equipped at the ET to achieve enormous beamforming gains; as a result, the end-to-end energy transfer efficiency can be greatly enhanced.

In practice, the benefit of energy beamforming in WET crucially depends on the available  channel state information (CSI) at the ET, which needs to be acquired at the cost of additional resources (e.g., time and energy). Similar to wireless communication systems, one straightforward approach to obtain CSI at the ET is by sending pilot signals from the ET to the ER \cite{495,500,547}, based on which the ER estimates the channel and sends the estimation back  to the ET via a feedback link. However, since the training time required scales up with the number of  antennas $M$ at the ET, such a channel-acquisition method is not suitable when $M$ becomes large \cite{484}. Furthermore, as pointed out by \cite{491}, estimating  the channel at the ER requires complex baseband signal processing, which may not be available  at the ER due to its practical hardware limitation. A novel channel-learning algorithm for WET by taking into account the practical energy harvesting circuitry at the ER has thus been proposed in \cite{491}, which simplifies the processing at the ER and requires only one-bit feedback from the ER per feedback interval. Nevertheless, the number of feedback intervals required for the channel-learning scheme in \cite{491} still increases quickly with $M$, which may be prohibitive for large $M$. Furthermore, it is worth noting that for all the aforementioned schemes, feedback signals need to be sent from the ER to the ET using part of the harvested energy. Considering the limited energy availability at the ER, the energy consumed for sending the feedback signals should be taken into account in the channel learning and feedback designs for MIMO WET.

 In this paper, we study the important channel acquisition problem for a point-to-point MIMO WET system in Rician fading channels, which are general enough to include a class of MIMO channels ranging from a fully random Rayleigh fading channel to a fully deterministic LOS channel by varying the Rician factor, $K$. Note that although only point-to-point MIMO system is considered in this paper for ease of exposition, the techniques presented are also applicable to the multiuser setup with simultaneous energy transmission to a group of single- or multi-antenna ERs with similar path losses, i.e., with roughly the same distance from the ET. We assume that the forward-link energy transmission from the ET to the ER and the reverse-link communication from the ER to the ET take place in a time-division duplexing (TDD) manner, so that channel reciprocity holds, i.e., the channel matrices in the forward and reverse links are transpose of each other. Notice  that channel reciprocity is a key enabling factor  for the implementation of massive MIMO systems to reduce the channel-acquisition overhead for next generation wireless communication systems \cite{373}.
 Under channel reciprocity, it is a well-known technique in wireless communication that the CSI  at the transmitter can be obtained via reverse-link channel training, where a fraction of the channel coherence time is assigned to the receiver for sending pilot signals to the transmitter to estimate the channel. However, applying this method to WET systems needs a more careful design of  the training strategy, since the pilot signals need to be sent using part of the energy harvested by the ER. In particular, the following new trade-offs need to be taken into account for the training design in WET systems: too little training leads to coarsely estimated channel at the ET and hence reduced  energy beamforming gain; whereas too much training consumes excessive energy harvested by the ER, and also leaves less time for energy transmission  given a finite channel coherence time. To resolve the above trade-offs,  we propose to maximize the  {\it net} harvested energy at the ER, which is the average harvested energy  offset by that used for channel training.  Specifically, the main contributions of this paper are summarized as follows:

 First, we propose a two-phase protocol for MIMO WET by leveraging the channel reciprocity property. In the first phase of $\tau$ symbol durations, pilot signals are sent from the ER to ET for channel training. Based on the received pilot signals, the ET obtains an estimate of the instantaneous MIMO channel. During the second phase of $T-\tau$ symbol durations where $T$ denotes the channel coherence time, the ET transmits energy-bearing signals with energy beamforming based on the estimated channel. Compared with the existing feedback-based channel-acquisition schemes considered in \cite{495,500,491}, the proposed channel-reciprocity enabled method has the following advantages: (i) it is more efficient for large or massive MIMO system implementation as the training overhead is independent of the number of antennas $M$ at the ET; (ii) it simplifies the processing at the ER since channel estimation and feedback operations are no longer required; and (iii) it explicitly maximizes the \emph{net} average  harvested energy  at the ER, which takes into account the time and energy overhead required for channel training.

   Second, for the general uncorrelated MIMO Rician fading channels with the proposed two-phase protocol, an optimization problem is formulated to maximize the net average harvested energy at the ER to obtain the optimal training design, which includes the optimal subset of ER antennas to be trained, as well as the training time and power that are allocated.

   Third, as finding the optimal solution to the training design problem is challenging in general, we focus on three special scenarios of high  practical interests to reveal useful  insights, namely, MIMO Rayleigh fading channels as well as MISO and massive MIMO Rician fading channels. For the former two cases, the optimal training designs are obtained via convex optimization techniques. For massive MIMO setup, an approximate solution for the training design problem is obtained based on a newly derived lower bound of the net average harvested energy. For the special case of rank-1 deterministic MIMO Rician channel component, we show that the obtained approximate solution achieves the optimal asymptotic performance  scaling as the ideal case  assuming perfect CSI at the ET as $M\rightarrow \infty$.

The rest of this paper is organized as follows. Section~\ref{sec:systemModel} introduces the system model. Section~\ref{sec:protocol} discusses the proposed two-phase protocol for MIMO WET and formulates the training optimization problem for MIMO Rician fading channels. In Section~\ref{sec:optimalTraining}, the formulated problem is solved for three special scenarios of MIMO Rayleigh fading channels, MISO and massive MIMO Rician fading channels, respectively. In Section~\ref{sec:numerical}, numerical results are provided to corroborate our study. Finally, we conclude the paper and point out some  future working directions in Section~\ref{sec:conclusion}.

\emph{Notations:} In this paper, scalars are denoted by italic letters. Boldface lower- and upper-case letters denote vectors and matrices, respectively. $\mathbb{C}^{M\times N}$ denotes the space of $M\times N$ complex matrices. $\mathbb{E}_{\mathbf X}[\cdot]$ denotes the expectation with respect to the random matrix $\mathbf X$.  $\mathbf{I}_M$ denotes an $M\times M$ identity matrix and $\mathbf{0}$ denotes an all-zero matrix. For an arbitrary-size matrix $\mathbf{A}$,  its transpose, Hermitian transpose, and Frobenius  norm are respectively denoted as $\mathbf{A}^{T}$, $\mathbf{A}^{H}$ and $\|\mathbf{A}\|_F$. For a square Hermitian matrix $\mathbf{S}$, $\mathrm{Tr}(\mathbf{S})$ denotes its trace, while $\lambda_{\max}(\mathbf S)$ and $\mathbf v_{\max}(\mathbf S)$ denote its largest eigenvalue and the corresponding eigenvector, respectively.   The symbol $j$ represents the imaginary unit of complex number, i.e., $j^2=-1$. $[\mathbf A]_{nm}$ denotes the $(n,m)$-th element of matrix $\mathbf A$, and $[\mathbf v]_{1:N_1}$ denotes a vector consisting of the first $N_1$ elements of vector $\mathbf v$. For a set $\mathcal{N}$, $|\mathcal{N}|$ denotes its cardinality. Furthermore, $\mathcal{N}\setminus \mathcal{N}_1$ denotes the complement of set $\mathcal N_1$ in $\mathcal{N}$.

 \section{System Model}\label{sec:systemModel}
\begin{figure*}
\centering
\includegraphics[scale=0.65]{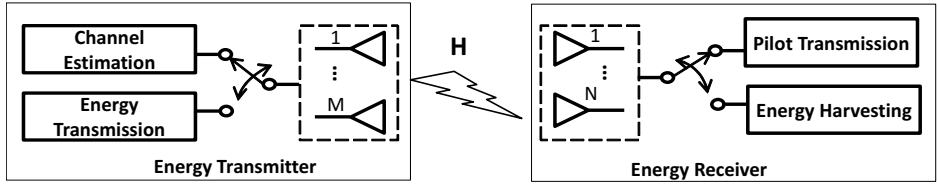}
\caption{A point-to-point MIMO WET system with reverse-link channel training and forward-link energy transmission.}\label{F:MIMOWET}
\end{figure*}

We consider a point-to-point MIMO WET system as shown in Fig.~\ref{F:MIMOWET}, where an ET with $M$ antennas is employed to deliver wireless energy to the ER, which is equipped with $N$ antennas. We assume a quasi-static flat fading channel model, where the baseband equivalent channel $\mathbf H\in \mathbb{C}^{N\times M}$ from the ET to the ER remains constant within each coherent block of $T$ symbol durations, and varies independently from one block to another. We consider the scenarios where a LOS link is generally  present between the ET and the ER, for which $\mathbf H$ can be modeled by the MIMO Rician fading channel as \cite{499}
\begin{align}
\mathbf H = \sqrt{\frac{\beta K}{K+1}}\Hbar + \sqrt{\frac{\beta}{K+1}}\Hw,\label{eq:HRician}
\end{align}
where $\Hbar$ is a deterministic $N\times M$ matrix containing the Rician (including the LOS) components of the channel, $\Hw$ is a random $N\times M$ matrix with independent and identically distributed (i.i.d.) zero-mean unit-variance circularly symmetric complex Gaussian (CSCG) entries, i.e., $[\Hw]_{nm}\sim \mathcal{CN}(0,1)$, which represent the scattered components of the channel, and $K\in [0, \infty)$ is the Rician $K$-factor denoting the power ratio between the Rician and the scattered components. Furthermore, the parameter $\beta$ models the large-scale fading, which includes the effects of both distance-dependent path loss and shadowing \cite{499}. Note that the deterministic Rician channel component $\Hbar$ can be modeled as \cite{519}
\begin{align}
\Hbar=\sum_{l=1}^L g_l \mathbf a_r(\theta_{r,l}) \mathbf a_t(\theta_{t,l})^H, \label{eq:Hbar}
\end{align}
where $L$ is the number of deterministic paths, $g_l$ is the complex amplitude of the $l$th path, and $\theta_{r,l}$ and $\theta_{t,l}$ are the angle of arrival (AOA) and angle of departure (AOD), respectively. Moreover,
$\mathbf a_r(\theta)$ and $\mathbf a_t(\theta)$ are the array responses at the ER and ET, respectively, which are given by \cite{440} 
\begin{equation}\label{eq:ar}
\begin{aligned}
\mathbf a_r(\theta)&=\left[\begin{matrix} 1, & e^{j\Phi_1(\theta)}, & \cdots, & e^{j\Phi_{(N-1)}(\theta)}\end{matrix} \right]^T \\
\mathbf a_t(\theta)&=\left[\begin{matrix} 1, & e^{j\Phi_1(\theta)}, & \cdots, & e^{j\Phi_{(M-1)}(\theta)} \end{matrix} \right]^T,
\end{aligned}
\end{equation}
where $\Phi_m$, $m=1,\cdots,N-1$ or $1,\cdots, M-1$, is the phase shift of the $m$th array element with respect to the reference antenna, which depends on the array configuration and is a function of the AOA/AOD. For the simple uniform linear array (ULA) configuration with antenna separation $d$ in wavelengths, we have
\begin{align}\label{eq:Phim}
\Phi_m(\theta)=2\pi m d \sin(\theta), \ \forall m.
\end{align}

 The total average received power of the channel given in \eqref{eq:HRician} can be expressed as
\begin{align}
\xE\left[\|\mathbf H\|_F^2\right]&=\frac{\beta K}{K+1}\tr(\Hbar \Hbar^H)+\frac{\beta}{K+1}\tr \left( \xE \left[\Hw\Hw^H\right]\right) \notag \\
&=\frac{\beta K}{K+1}\tr(\Hbar \Hbar^H) + \frac{\beta}{K+1}MN, \label{eq:HnormEq2}
\end{align}
where \eqref{eq:HnormEq2} is true since $\xE\left[\Hw\Hw^H\right]=M\mathbf I_N$.  Note that in the above expression, $\tr(\Hbar \Hbar^H)$ gives the total power gain of the deterministic channel components.  Without loss of generality, we assume
\begin{align}
 \tr(\Hbar \Hbar^H)=MN,\label{eq:trHbar}
\end{align}
 so that $\xE\left[\|\mathbf H\|_F^2\right]=\beta MN$, $\forall K\in [0, \infty)$. Therefore, $\beta$ can be interpreted as the average channel power per transmit-receive antenna pair.

 Note that by varying the Rician $K$-factor, the model in \eqref{eq:HRician}  includes a large class of MIMO wireless channels ranging from a fully random Rayleigh fading MIMO channel ($K=0$) to a fully deterministic MIMO channel ($K\rightarrow \infty$).

Within one coherent block of $T$ symbols, the input-output relation for the forward link of MIMO WET system can be written as
\begin{align}
\mathbf y[t]= \mathbf H \mathbf x[t] + \mathbf n[t], \ t=1,\cdots, T,\label{eq:yt}
\end{align}
where $\mathbf y[t]\in \mathbb{C}^{N\times 1}$ is the received signal vector at the ER in the $t$th channel use; $\mathbf n[t]$ denotes the additive Gaussian noise vector at the ER; and $\mathbf x[t]\in \mathbb{C}^{M\times 1}$, $t=1,\cdots, T$, represents the energy-bearing signal transmitted by the ET with \emph{sample} covariance matrix denoted by $\mathbf S$, i.e.,
\begin{align}\label{eq:sampleCov}
\mathbf S\triangleq \frac{1}{T}\sum_{t=1}^T \mathbf x[t] \mathbf x[t]^H.
 \end{align}
 Note that if $\{\mathbf x[t]\}_{t=1}^T$ are i.i.d. zero-mean random vectors generated from certain distributions, the sample covariance matrix $\mathbf S$ approaches to the statistical covariance matrix as $T\rightarrow \infty$. In contrast to that in wireless communications where the end-to-end information rate achievable depends on the complete probability distribution (including the transmit covariance matrix) of the transmitted signal vector, it will become clear later (cf. \eqref{eq:Q}) that only the sample covariance matrix $\mathbf S$ given in \eqref{eq:sampleCov} affects the amount of energy harvested at the ER during each coherent block in MIMO WET systems. Denote by $P_f$ the average power constraint imposed at the ET in the forward link transmission. We then have $\tr(\mathbf S)\leq P_f$.

At the ER, the incident RF power captured by each of the $N$ receiving antennas is converted to usable direct current (DC) power by a device called rectifier, which generally consists of a Schottky diode and a lower-pass filter (LPF) \cite{514}. Due to the law of energy conservation, the total harvested RF-band energy, denoted by $Q$, from all receiving antennas at the ER during each coherent block  is proportional to that of the received baseband signal \cite{478}, i.e.,
 \begin{equation} \label{eq:Q}
 \begin{aligned}
 Q&=\sum_{t=1}^T \eta \big \|\mathbf H \mathbf x[t] \big \|^2 = \eta T \tr\left(\mathbf H \Big( \frac{1}{T} \sum_{t=1}^T \mathbf x[t] \mathbf x[t]^H\Big)\mathbf H^H\right)\\
 & =\eta T \tr \left(\mathbf H^H \mathbf H \mathbf S\right),
 \end{aligned}
 \end{equation}
 where $0<\eta\leq 1$ denotes the energy harvesting efficiency at the ER. Note that we have ignored the energy harvested from the background noise since it is generally negligible as compared to that harvested from the dedicated energy signals. For convenience, we assume unit symbol duration throughout the paper so that the terms energy and power are used interchangeably. It has been shown in \cite{478} that in the ideal case with perfect CSI $\mathbf H$ at the ET, the optimal covariance matrix that maximizes the harvested energy $Q$ is
  \begin{align}
  \mathbf S^{\star}=P_f \mathbf v_1 \mathbf v_1^H, \label{eq:SiStar}
  \end{align}
  where $\mathbf v_1=\mathbf v_{\max}\left(\mathbf H^H \mathbf H\right)$ denotes the eigenvector corresponding to the dominant eigenvalue of $\mathbf H^H \mathbf H$. The resulting maximum harvested energy is given by
  \begin{align}
  Q_{\max}=\eta T P_f \lambda_{\max}\left(\mathbf H^H \mathbf H\right).\label{eq:Qmaximum}
  \end{align}
  Since $\mathbf S^{\star}$ is a rank-one matrix, the transmitted energy-bearing signal $\mathbf x[t]$ satisfying \eqref{eq:sampleCov} can be obtained with beamforming, i.e., $\mathbf x[t]=\sqrt{P_f} \mathbf v_1s[t]$, $t=1,\cdots, T$, where $s[t]$ is an arbitrary random signal with zero mean and unit sample variance satisfying $1/T\sum_{t=1}^T |s[t]|^2=1$.

    Note that $Q_{\max}$ in \eqref{eq:Qmaximum} is generally a random variable  depending  on the channel realization $\mathbf H$. The maximum average harvested energy over all random channel realizations can be expressed as
\begin{align}\label{eq:Qbarmax}
\bar Q_{\max}&=\xE_{\mathbf H}[Q_{\max}]=\eta T P_f  \xE_{{\mathbf H}}\left[ \lambda_{\max}\left(\mathbf H^H \mathbf H\right)\right].
\end{align}
With $\mathbf H$ given by the Rician model in \eqref{eq:HRician},  the probability density function (pdf) of the maximum eigenvalue  $\lambda_{\max}\left({\mathbf H}^H {\mathbf H}\right)$ has been extensively studied in the literature (see e.g. \cite{498}), based on which the expectation $\xE_{\mathbf H}\left[\lambda_{\max}({\mathbf H}^H {\mathbf H})\right]$ can be numerically computed.

\begin{figure}
\centering
\includegraphics[scale=0.6]{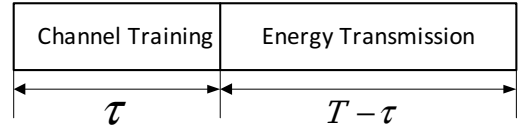}
\caption{A two-phase protocol for MIMO wireless energy transfer.}\label{F:twoPhase}
\end{figure}

In practice, only imperfect CSI can be made available at the ET due to the channel estimation and/or feedback errors. As a consequence, the maximum average harvested energy given in \eqref{eq:Qbarmax} cannot be achieved; instead, it only provides a performance upper bound for practical WET systems. In this paper, by exploiting the channel reciprocity between the forward (from the ET to the ER) and reverse (from the ER to the ET) links, we propose a two-phase protocol for the MIMO WET system for channel training and energy transmission, respectively. As illustrated in Fig.~\ref{F:twoPhase}, the first phase corresponds to the first $\tau\leq T$ symbol durations in each coherent block, where pilot symbols are sent by the ER to the ET  for channel training using the energy harvested in previous blocks. Based on the received pilot signals, the ET obtains an estimate of the MIMO channel. In the second phase of the remaining $T-\tau$ symbol durations, based on the estimated channel, the ET transmits the energy-bearing signal with optimized transmit beamforming. These two phases are elaborated in more details in the next section.

\section{Two-Phase Protocol}\label{sec:protocol}
 \subsection{Reverse-Link Channel Training}
 We assume that the channel parameters $\beta$, $K$, and $\Hbar$ are perfectly known at the ET since they typically vary slowly with time and hence are relatively easy to be estimated. On the other hand, the instantaneous channel realization $\mathbf H$, or equivalently $\Hw$, is estimated at the ET via reverse-link channel training. Since the entries in $\Hw$ are i.i.d. random variables, orthogonal pilot signals, which are known to be optimal \cite{2}, are assumed here to be sent by the ER. Denote by $\tau \leq T$ the number of symbol durations used for training in the first phase of each coherent block. To obtain an estimate for the complete channel matrix $\mathbf H$, we need at least as many measurements at the ET as unknowns, which implies that $M\tau\geq MN$ or $\tau \geq N$. Nevertheless, training for all ER antennas can be strictly suboptimal in general, as can be seen by considering the scenario with $T=N$, in which we have $\tau=T$ and hence no time is left for energy transmission in the second phase. Therefore, to obtain the optimal training design, we need to consider a more general strategy where possibly only a subset of the ER antennas is trained, which is denoted by the set $\mathcal{N}_1\subset \{1,\cdots, N\}$ with $|\mathcal{N}_1|=N_1$ and $0\leq N_1 \leq N$. Denote the corresponding channel matrix as $\Hone\in \mathbb{C}^{N_1\times M}$, i.e., $\Hone$ is a sub-matrix of $\mathbf H$ obtained by only choosing the rows with indices belonging to $\mathcal{N}_1$. Similarly, denote by $\mathcal{N}_2$ with $|\mathcal{N}_2|=N_2$ the set of non-trained ER antennas and $\Htwo\in \mathbb{C}^{N_2\times M}$ the corresponding channel matrix. We thus have $\mathcal{N}_2=\{1,\cdots, N\}\setminus \mathcal{N}_1$ and $N_2=N-N_1$. The channel matrix $\mathbf H$ can then be partitioned as
 \begin{align}
 \mathbf H=\boldsymbol \Pi \left[\begin{matrix}\Hone \\ \Htwo \end{matrix}\right], 
 \label{eq:HiDecomp}
 \end{align}
 where $\boldsymbol \Pi$ is an $N\times N$ permutation matrix that is determined by the choice of $\mathcal{N}_1$.

 With the Rician channel model given in  \eqref{eq:HRician}, the sub-matrices $\Hone$ and $\Htwo$ can be expressed as
 \begin{equation} \label{eq:H1and2}
\begin{aligned}
\Hone = \sqrt{\frac{\beta K}{K+1}} \Hbar_{\mathcal{N}_1} + \sqrt{\frac{\beta}{K+1}} \Hwone, \\
\Htwo = \sqrt{\frac{\beta K}{K+1}} \Hbar_{\mathcal{N}_2} + \sqrt{\frac{\beta}{K+1}} \Hwtwo,
\end{aligned}
\end{equation}
with $\Hbar_{\mathcal{N}_1}$, $\Hbar_{\mathcal{N}_2}$, $\Hwone$, and $\Hwtwo$ denoting the corresponding sub-blocks in $\Hbar$ and $\Hw$, respectively. During the training phase, pilot symbols are only sent from the $N_{1}$ ER antennas in $\mathcal{N}_1$, which yields
 \begin{align}
 \mathbf Y_{\mathrm{tr}}=\sqrt{\frac{P_r}{N_{1}}} \boldsymbol \Phi \Hone + \mathbf Z_{\mathrm{tr}}, \label{eq:Ytr}
 \end{align}
 where $\mathbf Y_{\mathrm{tr}}\in \mathbb{C}^{\tau \times M}$ contains the received training signals at the ET with $N_{1}\leq \tau \leq T$; $P_r$ is the training power used in the reverse link by the ER; $\boldsymbol \Phi\in \mathbb{C}^{\tau\times N_{1}}$ denotes the orthogonal pilot signals sent by the ER with $\boldsymbol \Phi^H \boldsymbol \Phi=\tau \mathbf I_{N_{1}}$; and $\mathbf Z_{\mathrm{tr}}\in \mathbb{C}^{\tau \times M}$ represents the noise received at the ET during the training phase with i.i.d. zero-mean CSCG entries each with variance $\sigma_r^2$. The energy consumed at the ER due to  channel training is given by 
 \begin{align}
 E_{\ttr}=\left\|\sqrt{\frac{P_r}{N_{1}}}\boldsymbol \Phi \right\|_F^2=\frac{P_r}{N_{1}}\tr(\boldsymbol \Phi^H \boldsymbol \Phi)=P_r\tau.\label{eq:Etr}
 \end{align}

 Based on  \eqref{eq:H1and2} and \eqref{eq:Ytr}, the minimum mean-square error (MMSE) estimate $\hat{\mathbf H}_{\mathcal{N}_1}$ of $\Hone$ can be expressed as \cite{71}
 \begin{align*}
 \hat{\mathbf H}_{\mathcal N_1} = \sqrt{\frac{\beta K}{K+1}} \Hbar_{\mathcal{N}_1} + \sqrt{\frac{\beta}{K+1}} \Hwonehat,
 \end{align*}
 where $\Hwonehat$ is the MMSE estimate of $\Hwone$ given by
 \begin{equation}
 \begin{aligned}
 \Hwonehat=& \frac{\sqrt{P_r \beta N_1 (K+1)}}{P_r \tau \beta + \sigma_r^2 N_1 (K+1)}  \boldsymbol \Phi^H \Bigg(\mathbf Y_{\mathrm{tr}}\\
 &-\sqrt{\frac{P_r \beta K}{N_1(K+1)}} \boldsymbol \Phi \Hbar_{\mathcal N_1} \Bigg).
 \end{aligned}
 \end{equation}
 Let $\Hwonetd$ denote the estimation error of $\Hwone$, i.e., $\Hwonetd=\Hwone-\Hwonehat$. Based on the well-known orthogonal property of the MMSE estimation for Gaussian random variables \cite{71}, $\Hwonehat$ and $\Hwonetd$ are independent. Furthermore, it can be obtained that $\Hwonehat$ and $\Hwonetd$ have i.i.d. zero-mean CSCG entries with variances  $\sigma^2_{\Hwonehat}$ and $\sigma^2_{\Hwonetd}$, respectively, where
 \begin{align}
 \sigma^2_{\Hwonehat}= \frac{\beta P_r \tau }{\beta P_r \tau+\sigma_r^2 N_1 (K+1)} \label{eq:sigmaH1hat}\\
 \sigma^2_{\Hwonetd}=\frac{\sigma_r^2 N_1(K+1)}{\beta P_r \tau + \sigma_r^2 N_1 (K+1)}\label{eq:sigmaH1td}.
 \end{align}

With \eqref{eq:HiDecomp}, \eqref{eq:H1and2} and the identity $\Hwone=\Hwonehat+\Hwonetd$, the channel matrix $\mathbf H$ can be decomposed as
\begin{equation}\label{eq:Hdecomp}
\begin{aligned}
\mathbf H = & \underbrace{\sqrt{\frac{\beta K}{K+1}} \boldsymbol \Pi \left[\begin{matrix} \Hbar_{\mathcal{N}_1} \\ \Hbar_{\mathcal{N}_2} \end{matrix} \right]
+ \sqrt{\frac{\beta}{K+1}} \boldsymbol \Pi \left[ \begin{matrix} \Hwonehat \\ \mathbf 0  \end{matrix} \right]}_{\triangleq \widehat{\mathbf H}}\\
&+\underbrace{\sqrt{\frac{\beta}{K+1}} \boldsymbol \Pi \left[ \begin{matrix} \Hwonetd \\ \Hwtwo \end{matrix}\right] }_{\triangleq \widetilde{\mathbf H}},
\end{aligned}
\end{equation}
where $\widehat{\mathbf H}$ denotes the CSI known at the ET, including the deterministic Rician complements $\Hbar$ as well as the estimated random components $\Hwonehat$ corresponding to the $N_1$ trained ER antennas in $\mathcal{N}_1$, and $\widetilde{\mathbf H}$ represents the CSI discrepancy from the true channel, including both the channel estimation error $\Hwonetd$ for the trained ER antennas and the un-estimated random components $\Hwtwo$ of the $N_2$ non-trained ER antennas. Note that since $\Hwonehat$, $\Hwonetd$, and $\Hwtwo$ are independent random matrices, $\widehat{\mathbf H}$ and $\widetilde{\mathbf H}$ are independent as well.

 \subsection{Forward-Link Energy Transmission}
 After the training-based channel estimation, energy-bearing signals are transmitted by the ET based on the estimated channel $\widehat{\mathbf H}$ during  the remaining $T-\tau$ symbol durations. Similar to \eqref{eq:Q},
 the total harvested energy  within one  coherent block can be expressed as
\begin{align}
Q&=\eta (T-\tau) \tr \left(\mathbf H^H \mathbf H \mathbf S\right).
\end{align}

Since the ET only has the knowledge of imperfectly and possibly partially estimated channel matrix $\widehat {\mathbf H}$ as given in \eqref{eq:Hdecomp}, the optimal transmit covariance matrix $\mathbf S^{\star}$ given in \eqref{eq:SiStar}  cannot be implemented.  In this case, given the knowledge of $\widehat {\mathbf H}$, $\mathbf S$ is optimized at the ET to maximize the average harvested energy $\widehat Q=\xE_{\widetilde{\mathbf H}}[Q|\widehat {\mathbf H}]$.
 With  the identity $\mathbf H=\widehat{\mathbf H}+\widetilde{\mathbf H}$, we have
\begin{align}
& \widehat {Q} = \eta (T-\tau) \xE_{\widetilde{\mathbf H}}\Big[\tr\big(\mathbf H^H \mathbf H \mathbf S\big) \big | \widehat{\mathbf H}\Big]  \notag \\
&=\eta (T-\tau)  \xE_{ \widetilde{\mathbf H}} \Big[\tr \big(\big(\widehat{\mathbf H}^H \widehat{\mathbf H}+\widetilde{\mathbf H}^H \widetilde{\mathbf H}+\widehat{\mathbf H}^H \widetilde{\mathbf H} + \widetilde{\mathbf H}^H \widehat{\mathbf H}\big)\mathbf S\big)\big|\widehat{\mathbf H}\Big] \notag  \\
&=\eta (T-\tau) \tr\left( \left( \widehat{\mathbf H}^H \widehat{\mathbf H}+\xE_{\widetilde{\mathbf H}}\left[\widetilde{\mathbf H}^H \widetilde{\mathbf H} \right]\right)\mathbf S \right) \label{eq:QhatEq1}\\
 &=\eta (T-\tau) \tr \Big(\Big(\widehat{\mathbf H}^H \widehat{\mathbf H} + \frac{\beta}{K+1}\big( N_1 \sigma^2_{\Hwonetd} + N_2 \big) \mathbf I_M \Big) \mathbf S\Big), \label{eq:QhatEq2}
\end{align}
where \eqref{eq:QhatEq1} follows from the independence between $\widetilde{\mathbf H}$ and $\widehat{\mathbf H}$, and \eqref{eq:QhatEq2} follows from the definition of $\widetilde{\mathbf H}$ given in \eqref{eq:Hdecomp}.
With the average transmit power constraint $P_f$ imposed at the ET, similar to \eqref{eq:SiStar}, the harvested energy $\widehat Q$ in \eqref{eq:QhatEq2} is maximized by setting the transmit covariance matrix to be
\begin{align}
\mathbf S &= P_f \mathbf v_E \mathbf v_E^H, \label{eq:SiEig}
\end{align}
where
\begin{align}
\mathbf v_E&=\mathbf v_{\max}\Big(\widehat{\mathbf H}^H \widehat{\mathbf H}+ \frac{\beta}{K+1}\big( N_1 \sigma^2_{\Hwonetd} + N_2 \big) \mathbf I_M\Big) \label{eq:SiImperfectCSI} \\
&=\mathbf v_{\max}\big(\widehat{\mathbf H}^H \widehat{\mathbf H}\big).\label{eq:SiImperfectCSI2}
\end{align}
The resulting maximum value of $\widehat Q$ is then given by
\begin{equation}\label{eq:Qstar}
\small
\begin{aligned}
\widehat Q^{\star}=\eta (T-\tau)P_f \left(\lambda_{\max} \big(\widehat{\mathbf H}^H \widehat{\mathbf H}\big)+ \frac{\beta}{K+1}\left( N_1 \sigma^2_{\Hwonetd} + N_2 \right)\right).
\end{aligned}
\end{equation}
It is observed  from \eqref{eq:SiImperfectCSI2} that, as the channel estimation error $\widetilde{\mathbf H}$ is isotropic,  the estimated channel matrix $\widehat{\mathbf H}$ can be treated as the true channel by the ET for maximum energy transfer.

With \eqref{eq:Qstar}, the average harvested energy $\bar{Q}$ can be expressed as
\begin{equation}\label{eq:QbarTraining3}
\small
\begin{aligned}
 \bar Q  (\mathcal{N}_1, \tau, P_r) =\xE_{\widehat{\mathbf H}}\big[\widehat Q^{\star}\big]=&\eta (T-\tau)P_f  \Big( \xE_{\Hwonehat} \Big[\lambda_{\max}\big(\widehat{\mathbf H}^H \widehat{\mathbf H}\big)\Big]\\
&+ \frac{\beta}{K+1}\left( N_1 \sigma^2_{\Hwonetd} + N_2 \right)\Big),
\end{aligned}
\end{equation}
where $\widehat {\mathbf H}$ is a random matrix dependent on $\Hwonehat$ given in \eqref{eq:Hdecomp}. Note that $\bar Q$ in \eqref{eq:QbarTraining3} depends on the subset of the training ER antennas $\mathcal{N}_1$, the training duration $\tau$, as well as the  training power $P_r$.

 The \emph{net} average harvested energy, i.e., the average harvested energy offset by  the energy consumed for sending pilot signals at the ER, is then given by 
\begin{align}\label{eq:Qnet}
\Qnet(\mathcal{N}_1, \tau, P_r)&=\bar Q (\mathcal{N}_1, \tau, P_r)-P_r \tau. 
\end{align}
 The problem of finding the optimal training design so that $\Qnet$ is maximized can thus be formulated as 
\begin{align}
\mathrm{(P1):} \qquad \underset{\mathcal{N}_{1}, \tau, P_r}{\max} \quad & \Qnet(\mathcal{N}_{1}, \tau, P_r) \notag \\
\text{s.t.} \quad & \mathcal{N}_1 \subset \{1,\cdots, N\}, \notag \\
& N_{1}\leq \tau \leq T, \notag \\
& P_r\geq 0. \notag
\end{align}
Note that there is an implicit constraint $P_r\tau < \bar Q (\mathcal{N}_1, \tau, P_r)$ in (P1) to ensure the strict positiveness of the objective value (net average energy). In (P1), we have assumed that there is a sufficiently large initial energy stored at the ER; under this assumption and further considering the fact that our optimized $\Qnet$ in (P1) is strictly positive and generally large enough (for implementing other functions at the ER such as sensing and data transmission), we can assume that there is always energy available at the ER to send the training signal with power $P_r$ at all time.

\section{Optimal Training Design}\label{sec:optimalTraining}
Finding the optimal solution to (P1) is challenging in general. One reason is that it involves discrete optimization over the set $\mathcal{N}_1$, which in general requires an exhaustive search over $2^N$ possibilities and thus becomes prohibitive for large $N$. Besides, even for fixed training set $\mathcal{N}_1$, finding the optimal solution for training duration $\tau$ and training power $P_r$ is non-trivial, which is mainly due to the difficulty in obtaining a closed-form expression for $\xE_{\Hwonehat} \Big[\lambda_{\max}\big(\widehat{\mathbf H}^H \widehat{\mathbf H}\big)\Big]$ for the general case. In this section, we derive the training strategies by efficiently solving (P1) in three special cases of high practical interests, namely, the MIMO Rayleigh fading channel, as well as the MISO and massive MIMO Rician fading channels.


\subsection{MIMO Rayleigh Fading Channel}\label{sec:Rayleigh}
We first consider the scenario where there is no LOS link between the ET and the ER. In this case, the MIMO Rician fading channel in \eqref{eq:HRician} reduces to MIMO Rayleigh fading channel with $K=0$. It then follows from the definition of $\widehat{\mathbf H}$ given in \eqref{eq:Hdecomp} that
\begin{align}
\xE_{\Hwonehat}\left[ \lambda_{\max}( \widehat{\mathbf H}^H \widehat{\mathbf H})\right]&=\beta \xE_{\Hwonehat} \left[ \lambda_{\max}(\Hwonehat^H\Hwonehat)\right]\\
&=\beta \sigma^2_{\Hwonehat} \xE_{\mathbf H_{\mathrm{n}}} \left[ \lambda_{\max}(\mathbf H_{\mathrm{n}}^H \mathbf H_{\mathrm{n}})\right]\\
&\triangleq \beta \sigma^2_{\Hwonehat} \Lambda(M,N_1),\label{eq:ExpRayleigh}
\end{align}
where $\mathbf H_{\mathrm{n}}\triangleq \Hwonehat/\sigma_{\Hwonehat}$ is the normalized channel matrix with i.i.d. zero-mean unit-variance CSCG entries. The pdf of the maximum eigenvalue  $\lambda_{\max}(\mathbf H_{\mathrm{n}}^H \mathbf H_{\mathrm{n}})$ has been extensively studied in the literature (see e.g. \cite{477}), based on which the expectation $\xE_{\mathbf H_{\mathrm{n}}} \left[ \lambda_{\max}(\mathbf H_{\mathrm{n}}^H \mathbf H_{\mathrm{n}})\right]$ can be computed. Note that $\xE_{\mathbf H_{\mathrm{n}}} \left[ \lambda_{\max}(\mathbf H_{\mathrm{n}}^H \mathbf H_{\mathrm{n}})\right]$  depends  only on the dimension of $\mathbf H_{\mathrm{n}}$ and hence is denoted as a function $\Lambda\left(M,N_1\right)$ in \eqref{eq:ExpRayleigh}. In the special cases of $N_1=1$ or $M=1$,  it can be easily obtained that
 $\Lambda\left(M,1\right)=M$ and $\Lambda\left(1,N_1\right)=N_1$. For general $M$ and $N_1$, no closed-form expression for $\Lambda\left(M,N_1\right)$ is available, whereas its numerical values can be easily computed, e.g., based on the algorithm proposed in \cite{476}. 


By substituting \eqref{eq:ExpRayleigh} into \eqref{eq:QbarTraining3} and applying \eqref{eq:sigmaH1hat} and \eqref{eq:sigmaH1td} with $K=0$, the average harvested energy for the WET system in MIMO Rayleigh fading channels can be expressed as
\begin{align}
\bar Q^{\mathrm{R}}(N_1,\tau,P_r)  = \eta (T-\tau)P_f \beta & \Big(\frac{ P_r \tau \beta \Lambda(M, N_{1})+\sigma_r^2 N_{1}^2}{P_r\tau\beta+\sigma_r^2N_{1}}\notag \\
&+N_2\Big). \label{eq:QbarFinalRayleigh}
\end{align}
Several observations can be made from \eqref{eq:QbarFinalRayleigh}.  First, the training antenna subset $\mathcal N_1$ affects the average harvested energy only through its cardinality $N_1$. 
 This is expected since for MIMO Rayleigh fading channels without the deterministic LOS component, all the $N$ ER antennas are statistically equivalent. Therefore, determining the optimal training antenna subset $\mathcal{N}_1$ is tantamount to finding the optimal number of training antennas $N_1$. Another observation from \eqref{eq:QbarFinalRayleigh} is that the average harvested energy $\bar Q^{\mathrm{R}}$ is given by a summation of two terms. The first term, which monotonically increases  with the training energy $P_r\tau$ and the number of ET antennas $M$, is attributed to the  $N_{1}$ trained ER antennas whose channel matrix $\Hone$ is estimated at the ET. The second term is attributed to the $N_{2}$ non-trained ER antennas, which is independent of the number of ET antennas $M$ since no beamforming gain can be achieved for the energy transmission over the associated channel $\Htwo$.

The net average harvested energy, as defined in \eqref{eq:Qnet}, can then be explicitly written as
\begin{equation}\label{eq:QnetRayleigh}
\begin{aligned}
\bar Q^{\mathrm{R}}_{\text{net}}(N_1,\tau,P_r)  = & \eta (T-\tau)P_f \beta \Bigg(\frac{ P_r \tau \beta \Lambda(M, N_{1})+\sigma_r^2 N_{1}^2}{P_r\tau\beta+\sigma_r^2N_{1}}\\
&+N-N_{1}\Bigg)-P_r\tau.
\end{aligned}
\end{equation}
Therefore, the training optimization problem (P1) for the case of MIMO Rayleigh fading channels can be formulated as
\begin{align}
\PRayleigh: \qquad \underset{N_{1}, \tau, P_r}{\max} \quad & \bar Q^{\mathrm{R}}_{\text{net}}(N_1,\tau,P_r) \notag \\
\text{s.t.} \quad & 0\leq N_{1} \leq N, \notag  \\
& N_{1}\leq \tau \leq T, \notag \\
& P_r\geq 0. \notag
\end{align}

To optimally solve problem $\PRayleigh$, we first show the following lemma.
\begin{lemma}\label{lemma:tauEqualN1}
The optimal solution $(N_1^{\star}, \tau^{\star}, P_r^{\star})$ to problem $\PRayleigh$ satisfies $\tau^{\star}=N_1^{\star}$.
\end{lemma}
\begin{IEEEproof}
Please refer to Appendix~\ref{A:tauEqualN1}.
\end{IEEEproof}

 By applying Lemma~\ref{lemma:tauEqualN1}, $\PRayleigh$ can be recast as
\begin{equation}\label{P2:Rayleigh}
\begin{aligned}
\qquad \underset{N_{1}, P_r}{\max} \quad & \bar{Q}_{\text{net}}^{\mathrm{R}}(N_{1}, P_r) \\
\text{s.t.} \quad & 0\leq N_{1} \leq N, \  P_r\geq 0,
\end{aligned}
\end{equation}
where $\bar{Q}_{\text{net}}^{\mathrm{R}}(N_{1}, P_r)$ is obtained by substituting $\tau=N_1$ into \eqref{eq:QnetRayleigh} and is given by
\begin{equation}\label{eq:Qnet2}
\begin{aligned}
\bar{Q}_{\text{net}}^{\mathrm{R}}(N_{1}, P_r)=&\eta (T-N_1)P_f \beta \Big(\frac{P_r \beta \Lambda(M, N_{1})+\sigma_r^2 N_{1}}{P_r\beta+\sigma_r^2}\\
&+N-N_{1}\Big)-P_rN_1.
\end{aligned}
\end{equation}

To find the optimal solution to problem \eqref{P2:Rayleigh}, we first obtain the optimal training power $P_r$ with $N_1$ fixed by solving the following optimization problem:
\begin{equation}\label{P3:Rayleigh}
 \begin{aligned}
\underset{P_r\geq 0}{\max} \quad & \bar{Q}_{\text{net}}^{\mathrm{R}}(N_{1}, P_r).
\end{aligned}
\end{equation}

Denote the optimal solution and the optimal value of problem \eqref{P3:Rayleigh} as $P_r^{\mathrm{R}}(N_1)$ and $\bar{Q}_{\text{net}}^{\mathrm{R}}(N_1)$, respectively. When $N_1=0$, it follows trivially that $P_r^{\mathrm{R}}(0)=0$ and the corresponding  energy is
\begin{align}
\bar{Q}_{\text{net}}^{\mathrm{R}}(0)=\eta T P_f \beta N. \label{eq:QnetN1Equal0}
\end{align}
For $1\leq N_1\leq N$, it can be verified that $\bar{Q}_{\text{net}}^{\mathrm{R}}(N_{1}, P_r)$ given in \eqref{eq:Qnet2} is a concave function with respect to $P_r$; hence problem \eqref{P3:Rayleigh} is  convex, whose solution can be found as
\begin{equation}\label{eq:PrstarN1}
\begin{aligned}
P_r^{\mathrm{R}}(N_1)=
\sqrt{\eta P_f \sigma_r^2}\Bigg[\sqrt{(T-N_1)\Big(\frac{\Lambda(M,N_1)}{N_1}-1\Big)}-\frac{1}{\sqrt{\Gamma}}\Bigg]^+,
\end{aligned}
\end{equation}
where $[x]^+\triangleq \max\{x,0\}$, and
\begin{align}\label{eq:Gamma}
\Gamma\triangleq  \eta P_f \beta^2/\sigma_r^2
 \end{align}
 is referred to as the two-way {\it effective} signal-to-noise ratio (ESNR). Note that the term $\beta^2$ in $\Gamma$ captures the effect of two-way signal attenuation  due to both the reverse-link training and the forward-link energy transmission.
The optimal value of problem \eqref{P3:Rayleigh} for $1\leq N_1\leq N$ can
then be expressed as \eqref{eq:QnetN1Greater0} given on  top of the next page,
\begin{figure*}
\begin{align}
\bar{Q}_{\text{net}}^{\mathrm{R}}(N_1)=
\begin{cases}
(T-N_1)\eta P_f \beta N +\eta P_f \beta N_1 \Bigg(\sqrt{(T-N_1)\Big(\frac{\Lambda(M,N_1)}{N_1}-1\Big)} -\frac{1}{\sqrt{\Gamma}}\Bigg)^2,  & \text{if } N_1 \in \mathcal{N}_{\mathrm{R}}\\
(T-N_1)\eta P_f \beta N, & \text{otherwise}, \label{eq:QnetN1Greater0}
\end{cases}
\end{align}
\hrulefill
\vspace{-2ex}
\end{figure*}
where the set $\mathcal N_{\mathrm{R}}$ is defined as
\begin{align}
\mathcal{N}_{\mathrm{R}}\triangleq \left \{1\leq N_1\leq N:  (T-N_1)\Big(\frac{\Lambda(M,N_1)}{N_1}-1\Big)> \frac{1}{\Gamma} \right\}.\label{eq:Nset}
\end{align}

Finding the optimal solution to problem \eqref{P2:Rayleigh}  and that to the original problem $\PRayleigh$ now reduces to determining the optimal number of ER antennas $N_1^{\star}$ to be trained, which can be easily obtained by comparing $\bar{Q}_{\text{net}}^{\mathrm{R}}(N_1)$ for the $N+1$ possible values of $N_1\in \{0,1,\cdots, N\}$. In fact, as evident from \eqref{eq:QnetN1Equal0} and \eqref{eq:QnetN1Greater0}, if $N_1\geq 1$ and $N_1 \notin \mathcal{N}_{\mathrm{R}}$, we always have $\bar{Q}_{\text{net}}^{\mathrm{R}}(N_1)<\bar{Q}_{\text{net}}^{\mathrm{R}}(0)$. Therefore, the searching space for $N_1^{\star}$ can be reduced as
\begin{align}
N_1^{\star}=\mathrm{arg} \underset{N_1\in \{0\} \cup \mathcal{N}_{\mathrm{R}} }{\max} \ \bar{Q}_{\text{net}}^{\mathrm{R}}(N_1), \label{eq:N1star}
\end{align}
which can be readily determined given the closed-form expressions \eqref{eq:QnetN1Equal0} and \eqref{eq:QnetN1Greater0}.

It follows from \eqref{eq:N1star} that $N_1^{\star}>0$ only when the set $\mathcal{N}_{\mathrm{R}}$ is non-empty. Thus, it can be inferred from \eqref{eq:Nset} that for WET system in MIMO Rayleigh fading channels, channel training helps only if the following conditions are satisfied: (i) the channel coherence time $T$ is sufficiently large; (ii) the ratio $\Lambda(M,N_1)/N_1$ is sufficiently large, i.e., the number of ET antennas $M$ is large enough; and (iii) the ESNR  $\Gamma$ is sufficiently high; otherwise, the benefit of channel training and hence the energy beamforming based on the estimated channel cannot compensate the time and the energy used for sending the pilot symbols by the ER, and thus no training should be applied; instead, it is optimal to assign all the $T$ symbol durations to transmit the energy signals isotropically (since there is no LOS channel in this case), as can be seen from \eqref{eq:SiImperfectCSI} that with $N_1=0$ and hence $\widehat{\mathbf H}=\mathbf 0$, the energy beamforming vector $\mathbf v_E$ can be set as any arbitrary unit-norm vector.

Note that for the special case of MISO setup with $N=1$, by applying the identity $\Lambda(M,1)=M$, the optimal solution to  \eqref{eq:N1star} can be found as
\begin{align}\label{eq:N1starRayleighMISO}
N_{1}^{\star}=
\begin{cases}
1, & \text{if } TM-T-M > \frac{1}{\Gamma}+\frac{2}{\sqrt{\Gamma}}\\
0, & \text{otherwise}.
\end{cases}
\end{align}

\subsection{MISO Rician Fading Channel}\label{sec:MISO}
In this subsection, we derive the optimal training strategy by solving (P1) for the case of MISO Rician fading channels, i.e., $N=1$ and $K>0$.  In this case, the antenna training set $\mathcal{N}_1$ in (P1)  has only two possibilities: $\mathcal{N}_1=\emptyset$ and $\mathcal{N}_1=\{1\}$, or equivalently, $N_1=0$ and $N_1=1$, i.e., only a binary decision needs to be made at the ER to send the pilot signals or not. Furthermore, the channel matrix $\mathbf H$ given in \eqref{eq:HRician} reduces to a $1\times M$ vector, and hence is denoted by a vector notation $\mathbf h$.\footnote{Similarly, all other matrix notations defined in the previous sections are replaced by the corresponding vector notations throughout this subsection when MISO channel is considered, e.g., $\hat {\mathbf h}$ and $\tilde{\mathbf h}$ correspond to $\widehat{\mathbf H}$ and $\widetilde{\mathbf H}$ defined in \eqref{eq:Hdecomp} with $N=1$, respectively.} The expectation in \eqref{eq:QbarTraining3} for both $N_1=0$ and $N_1=1$ can be explicitly expressed as
\begin{align}
\xE_{\hat{\mathbf h}_{\mathrm{w},\mathcal{N}_1}}\left[\lambda_{\max}(\hat{\mathbf h}^H \hat{\mathbf h}) \right]
&=\xE_{\hat{\mathbf h}_{\mathrm{w},\mathcal{N}_1}} \left[ \| \hat{\mathbf h} \|^2 \right] \notag \\
&=\frac{\beta K}{K+1} \|\bar{\mathbf h}\|^2+ \frac{\beta}{K+1} \xE\left[\|\hat{\mathbf h}_{\mathrm{w,1}}\|^2\right] \notag \\
&=\frac{\beta M K }{K+1} + \frac{\beta M N_1}{K+1}\sigma^2_{\hat{\mathbf h}_{\mathrm{w,1}}},\label{eq:MISOeq1}
\end{align}
where in \eqref{eq:MISOeq1} we have used the identity \eqref{eq:trHbar}. By substituting \eqref{eq:MISOeq1}  into \eqref{eq:QbarTraining3} and \eqref{eq:Qnet}, and applying  \eqref{eq:sigmaH1hat} and \eqref{eq:sigmaH1td}, the net average harvested energy for MISO Rician fading channels can be explicitly expressed as \eqref{eq:QnetMISO} given on top of the next page.
\begin{figure*}
\begin{align}
\bar{Q}_{\text{net}}^{\text{MISO}}(N_1,\tau,P_r)= & \eta (T-\tau) P_f \beta \Bigg(\frac{KM }{K+1}
+ \frac{N_1\left(M \beta P_r \tau/(K+1) +\sigma_r^2\right) }{\beta P_r \tau+\sigma_r^2 N_1 (K+1)}
+\frac{1-N_1}{K+1}\Bigg) - P_r \tau.\label{eq:QnetMISO}
\end{align}
\hrulefill
\end{figure*}

The training optimization problem (P1) for the case of MISO Rician fading channels can then be formulated as
\begin{align}
\PMISO: \qquad \underset{N_{1}, \tau, P_r}{\max} \quad & \bar{Q}_{\text{net}}^{\text{MISO}}(N_1,\tau,P_r) \notag \\
\text{s.t.} \quad & N_{1}\in \{0, 1\}, \notag \\
& N_{1}\leq \tau \leq T, \notag \\
& P_r\geq 0. \notag
\end{align}

The optimal solution to $\PMISO$ can be obtained by separately considering the two cases with $N_1=0$ and $N_1=1$.

{\it Case 1: $N_1=0$}, i.e. no channel training is performed. In this case, \eqref{eq:QnetMISO} reduces to
  \begin{align}
  \bar{Q}_{\text{net}}^{\text{MISO}}(0,\tau,P_r)=\eta (T-\tau) P_f \beta \frac{KM+1}{K+1}-P_r\tau.
  \end{align}
  It then follows trivially  that we should have $\tau=0$ and $P_r=0$, which yields
\begin{align}
\bar{Q}_{\text{net}}^{\text{MISO}}(N_1=0)=\eta T P_f \beta \frac{KM+1}{K+1}.\label{eq:QnetMISONoTrain}
\end{align}
It can be obtained based on \eqref{eq:SiImperfectCSI} and \eqref{eq:Hdecomp} that the average harvested energy given in \eqref{eq:QnetMISONoTrain} with no channel training is achieved by the ET sending the energy-bearing signals only based on the LOS component of the channel, which is intuitively understood since the random channel component is isotropic and is not estimated at the ET.

{\it Case 2: $N_1=1$}. In this case, (P1-MISO) reduces to
\begin{equation}\label{P:MISON1equal1}
\begin{aligned}
\underset{\tau, P_r}{\max} \quad & \eta (T-\tau) P_f \beta \Bigg(\frac{KM }{K+1}
+ \frac{M \beta P_r \tau/(K+1) +\sigma_r^2}{\beta P_r \tau+\sigma_r^2 (K+1)}
\Bigg) - P_r \tau  \\
\text{s.t.} \quad & 1\leq \tau \leq T, \\
& P_r\geq 0.
\end{aligned}
\end{equation}

Following similar derivations as in the previous subsection, it can be obtained that the optimal solution $(\tau^{\star}, P_r^{\star})$ to problem \eqref{P:MISON1equal1} is
\begin{align}\label{eq:optSolMISO}
\tau^{\star}=1, \ P_r^{\star}=\sqrt{\eta P_f \sigma_r^2}\left[\sqrt{(T-1)(M-1)}-\frac{K+1}{\sqrt{\Gamma}} \right]^+,
\end{align}
where $\Gamma$ is the ESNR defined in \eqref{eq:Gamma}. The corresponding optimal value of \eqref{P:MISON1equal1} is
\begin{align}\label{eq:QnetMISOWithTrain}
\bar{Q}_{\text{net}}^{\text{MISO}}(1)=
\begin{cases}
(T-1)\eta P_f \beta \frac{KM+1}{K+1}+ \frac{\eta P_f \beta}{K+1} \bigg(\sqrt{(T-1)(M-1)}\\
\hspace{1ex} -\frac{K+1}{\sqrt{\Gamma}} \bigg)^2, & \hspace{-30ex} \text{ if } (T-1)(M-1) > \frac{(K+1)^2}{\Gamma} \\
(T-1)\eta P_f \beta \frac{KM+1}{K+1},   & \hspace{-25ex} \text {otherwise}.
\end{cases}
\end{align}


The optimal solution to $\PMISO$ can then be obtained by comparing $\bar{Q}_{\text{net}}^{\text{MISO}}(N_1=0)$ in \eqref{eq:QnetMISONoTrain} and $\bar{Q}_{\text{net}}^{\text{MISO}}(N_1=1)$ in \eqref{eq:QnetMISOWithTrain}. After some further derivations, the optimal value for $N_1$ to problem $\PMISO$ can be obtained as
\begin{align}\label{eq:N1OptMISO}
N_1^{\star}=
\begin{cases}
1, & \text{ if } (T-1)(M-1) > \left( \sqrt{KM+1} + \frac{K+1}{\sqrt{\Gamma}}\right)^2\\
0, & \text{ otherwise.}
\end{cases}
\end{align}

It is observed from \eqref{eq:N1OptMISO} that, similar to the case of MIMO Rayleigh fading channels, channel training helps for WET in MISO Rician fading channels only if $T$, $M$, and $\Gamma$ are sufficiently large. In addition, the Rician factor $K$ needs to be sufficiently small for channel training to be effective; otherwise, it is optimal for the ET to send with the energy beamforming based on the deterministic Rician components $\bar {\mathbf h}$ of the channel.

Note that by setting $K=0$, the solution \eqref{eq:N1OptMISO} reduces to that given in \eqref{eq:N1starRayleighMISO} for the special case of MISO Rayleigh fading channel.

\subsection{Massive MIMO Rician Fading Channel}\label{sec:massive}
Massive MIMO techniques have been recently advanced to tremendously improve the energy and spectrum efficiency of wireless communication networks by deploying a very large number of antennas (say, hundreds or even more) at the base stations \cite{374,497}.  For WET system with a large number of antennas deployed at the ET, it is expected that the end-to-end energy transfer efficiency can be similarly enhanced, provided that the channel is properly learned at the ET to achieve the enormous beamforming gain. In this subsection, we optimize the training design for massive MIMO WET systems by solving the optimization problem (P1) with a large value of $M$, i.e., $M\gg N$.

 Based on the definition of $\widehat{\mathbf H}$ in \eqref{eq:Hdecomp} and the law of large numbers, the following asymptotic result holds for large $M$:
\begin{equation}\label{eq:massiveMIMOApprox}
\small
\begin{aligned}
\hspace{-3ex} \frac{1}{M} \widehat{\mathbf H}\widehat{\mathbf H}^H &\overset{a.s.}{\rightarrow} \frac{1}{M} \frac{\beta K}{K+1} \Hbar \Hbar^H + \frac{\beta}{K+1} \boldsymbol \Pi \left[\begin{matrix} \frac{1}{M} \Hwonehat \Hwonehat^H & \mathbf 0 \\ \mathbf 0 & \mathbf 0\end{matrix}\right] \boldsymbol \Pi^T \\
&\overset{a.s.}{\rightarrow} \frac{1}{M} \frac{\beta K}{K+1} \Hbar \Hbar^H + \frac{\beta \sigma^2_{\Hwonehat}}{K+1} \boldsymbol \Pi \left[\begin{matrix} \mathbf I_{N_1} & \mathbf 0 \\ \mathbf 0 & \mathbf 0\end{matrix}\right] \boldsymbol \Pi^T,
\end{aligned}
\end{equation}
 where $\overset{a.s.}{\rightarrow}$ denotes the almost sure convergence.
As a result, we have
\begin{align}
\hspace{-3ex} &\xE_{\hat{\mathbf H}_{\mathrm{w},\mathcal{N}_1}}\left[\lambda_{\max}(\widehat{\mathbf H}^H\widehat{\mathbf H})\right]
=\xE_{\hat{\mathbf H}_{\mathrm{w},\mathcal{N}_1}}\left[\lambda_{\max}(\widehat{\mathbf H}\widehat{\mathbf H}^H)\right] \label{eq:avgMaxEig}\\
&\overset{a.s.}{\rightarrow} \lambda_{\max} \left(\frac{\beta K}{K+1} \Hbar \Hbar^H + \frac{ \beta M \sigma^2_{\Hwonehat}}{K+1} \boldsymbol \Pi \left[\begin{matrix} \mathbf I_{N_1} & \mathbf 0 \\ \mathbf 0 & \mathbf 0\end{matrix}\right] \boldsymbol \Pi^T
\ \right) \notag \\
& \geq \frac{\beta K \bar{\lambda}}{K+1} + \frac{\beta M \sigma^2_{\Hwonehat}}{K+1} \left \|\big[ \boldsymbol \Pi^T \bar{\mathbf v}\big]_{1:N_1}\right \|^2, \label{eq:LBLambdaMax}
\end{align}
where $\bar{\lambda}$ and $\bar{\mathbf v}$ are the dominant eigenvalue and the corresponding eigenvector of the matrix $\Hbar \Hbar^H$, respectively, i.e., $\bar{\lambda}\triangleq \lambda_{\max}(\Hbar \Hbar^H)$, and $\bar{\mathbf v}\triangleq \mathbf v_{\max} ( \Hbar \Hbar^H)$. The inequality in \eqref{eq:LBLambdaMax} is obtained by directly applying the following result for any two Hermitian matrices $\mathbf A$ and $\mathbf B$:
\begin{align}
\lambda_{\max}(\mathbf A + \mathbf B) = \underset{\|\mathbf v\|=1}{\max}\  \mathbf v^H (\mathbf A + \mathbf B) \mathbf v \\
\geq \mathbf v_{\mathbf A}^H \mathbf A \mathbf v_{\mathbf A} + \mathbf v_{\mathbf A}^H \mathbf B \mathbf v_{\mathbf A} \label{eq:inequ}\\
=\lambda_{\max}(\mathbf A) + \mathbf v_{\mathbf A}^H \mathbf B \mathbf v_{\mathbf A},
\end{align}
where $\mathbf v_{\mathbf A}$ is the eigenvector corresponding to the dominant eigenvalue of matrix $\mathbf A$. Note that equality holds in \eqref{eq:inequ} when $\mathbf A$ and $\mathbf B$ have the same dominant eigenvector.

By applying \eqref{eq:sigmaH1hat}, \eqref{eq:sigmaH1td}, and \eqref{eq:LBLambdaMax} into \eqref{eq:Qnet}, it can be obtained that the net average harvested energy for massive MIMO Rician fading channels with large $M$ is lower-bounded as \eqref{eq:QnetLB} given on top of the next page.
\begin{figure*}
\begin{align}
\Qnet(\mathcal N_1, \tau, P_r) & \geq
\frac{\eta(T-\tau)P_f \beta}{K+1} \bigg(K \bar{\lambda} +
 \frac{M \left\|\big[ \boldsymbol \Pi^T \bar{\mathbf v}\big]_{1:N_1}\right\|^2 \beta P_r \tau + N_1^2 \sigma_r^2 (K+1)}{\beta P_r \tau+ N_1 \sigma_r^2(K+1)}  + N-N_1\bigg)-P_r\tau \label{eq:QnetLB} \\
&\triangleq \tilde{Q}_{\text{net}}^{\text{large-}M}(\boldsymbol \Pi, N_1,\tau,P_r).
\end{align}
\hrulefill
\end{figure*}

 Based on \eqref{eq:LBLambdaMax} and \eqref{eq:inequ}, it can be shown  that the above lower bound is tight when $N_1=0$ or $N_1=N$.  Note that in \eqref{eq:QnetLB}, the training antenna subset $\mathcal N_1$ affects $\tilde{Q}_{\text{net}}^{\text{large-}M}$ via both the corresponding permutation matrix $\boldsymbol \Pi$ and the cardinality $N_1$.  In fact, as can be seen from \eqref{eq:HiDecomp}, $\mathcal N_1$ can be determined once $\boldsymbol \Pi$ and $N_1$ are obtained, by re-ordering all the $N$ ER antennas based on the permutation matrix $\boldsymbol \Pi^T$ and then selecting the first $N_1$ elements as the set $\mathcal N_1$.  Therefore, the training optimization problem (P1) in massive MIMO Rician fading channels can be approximately solved by maximizing the lower bound given in \eqref{eq:QnetLB} as
\begin{equation*}
\begin{aligned}
\PMassive: \quad \underset{\boldsymbol \Pi, N_{1}, \tau, P_r}{\max} \quad & \tilde{Q}_{\text{net}}^{\text{large-}M}(\boldsymbol \Pi, N_{1}, \tau, P_r) \\
\text{s.t.}: \quad & \boldsymbol \Pi \text{ is a permutation matrix} \\
& 0\leq N_{1} \leq N, \\
& N_{1}\leq \tau \leq T, \\
& P_r\geq 0.
\end{aligned}
\end{equation*}

To find the optimal solution to $\PMassive$, we first determine the optimal permutation matrix $\boldsymbol \Pi$. It can be inferred from \eqref{eq:QnetLB} that for any given $N_1\in \{0,1,\cdots,N\}$, $\tilde{Q}_{\text{net}}^{\text{large-}M}$ is maximized if the absolute values of the elements in the vector $\boldsymbol \Pi^T \bar{\mathbf v}$ are arranged in a non-decreasing order, which thus gives the optimal permutation matrix $\boldsymbol \Pi^{\star}$ for $\PMassive$. 
Furthermore, with similar arguments as that for the proof of Lemma~\ref{lemma:tauEqualN1}, it can be verified that the optimal training duration $\tau^\star$ should be equal to the optimal number of antennas $N_1^\star$ to be trained. As a result, $\PMassive$ reduces to
\begin{equation}\label{P:massiveMIMOP2}
\begin{aligned}
\qquad \underset{N_{1}, P_r}{\max} \quad & \tilde{Q}_{\text{net}}^{\text{large-}M}(N_{1}, P_r) \\
\text{s.t.} \quad & 0\leq N_{1} \leq N, \\
& P_r\geq 0,
\end{aligned}
\end{equation}
where $\tilde{Q}_{\text{net}}^{\text{large-}M}(N_{1}, P_r)$ is obtained by substituting $\boldsymbol \Pi=\boldsymbol \Pi^{\star}$ and $\tau=N_1$ into \eqref{eq:QnetLB}, which is
\begin{equation}
\begin{aligned}
&\tilde{Q}_{\text{net}}^{\text{large-}M}(N_{1}, P_r)=
\frac{\eta(T-N_1)P_f \beta}{K+1} \bigg(K \bar{\lambda} \\
&\hspace{1ex} +
 \frac{M \|[\bar{\mathbf v}^{\star}]_{1:N_1}\|^2 \beta P_r  + N_1 \sigma_r^2 (K+1)}{\beta P_r +  \sigma_r^2(K+1)}  + N-N_1\bigg)-P_r N_1,
\end{aligned}
\end{equation}
with $\bar{\mathbf v}^{\star}\triangleq \boldsymbol \Pi^{\star T} \bar{\mathbf v}$ obtained by rearranging the elements in $\bar{\mathbf v}$ so that the elements' absolute values are in a non-increasing order.

To find the optimal solution to problem \eqref{P:massiveMIMOP2}, we first obtain the optimal training power with $N_1$ fixed by solving the following optimization problem:
\begin{equation}\label{P:massiveMIMOP3}
\begin{aligned}
\underset{P_r\geq 0}{\max} \quad \tilde{Q}_{\text{net}}^{\text{large-}M}(N_{1}, P_r).
\end{aligned}
\end{equation}
Denote the optimal solution and the optimal value of problem \eqref{P:massiveMIMOP3} as $P_r^{\text{large-}M}(N_1)$ and $\tilde{Q}_{\text{net}}^{\text{large-}M}(N_1)$, respectively. When $N_1=0$, it follows trivially that $P_r^{\text{large-}M}(0)=0$ and the objective value is
\begin{align}\label{eq:QnetMassiveNoTrain}
\tilde{Q}_{\text{net}}^{\text{large-}M}(0)=\eta T P_f \beta \frac{K \bar{\lambda} +N}{K+1}.
\end{align}

For $1\leq N_1 \leq N$, it can be verified that problem \eqref{P:massiveMIMOP3} is convex, whose optimal solution is given by
\begin{equation}
\begin{aligned}
\hspace{-3ex} P_r^{\text{large-}M}(N_1)=& \sqrt{\eta P_f \sigma_r^2} \bigg[\sqrt{(T-N_1)\Big(\frac{M \|[\bar{\mathbf v}^\star]_{1:N_1}\|^2}{N_1}-1\Big)}\\
& -\frac{K+1}{\sqrt{\Gamma}} \bigg]^+,
\end{aligned}
\end{equation}
and the corresponding optimal value can be expressed as \eqref{eq:QnetMassiveWithTrain} given on top of the next page,
\begin{figure*}
\begin{align}\label{eq:QnetMassiveWithTrain}
\tilde{Q}^{\text{large-}M}_{\text{net}}(N_1)=
\begin{cases}
(T-N_1)\eta P_f \beta \frac{K \bar{\lambda}+N}{K+1} + \frac{\eta P_f \beta N_1}{K+1} \left(\sqrt{(T-N_1)\left(\frac{M \|[\bar{\mathbf v}^\star]_{1:N_1}\|^2}{N_1}-1\right)}-\frac{K+1}{\sqrt{\Gamma}} \right)^2, & \text{ if } N_1 \in \mathcal{N}_{\text{large-}M} \\
(T-N_1)\eta P_f \beta \frac{K \bar{\lambda}+N}{K+1}, &  \text{ otherwise}
\end{cases}
\end{align}
\hrulefill
\vspace{-2ex}
\end{figure*}
where the set $\mathcal{N}_{\text{large-}M}$ is defined as
\begin{equation}\label{eq:setNMassive}
\begin{aligned}
\hspace{-2ex} \mathcal{N}_{\text{large-}M}\triangleq \Bigg\{1\leq N_1 \leq N: & (T-N_1)\left(\frac{M \|[\bar{\mathbf v}^\star]_{1:N_1}\|^2}{N_1}-1\right)\\
& >\frac{(K+1)^2}{\Gamma} \Bigg\}.
\end{aligned}
\end{equation}

Finding the optimal solution to problem \eqref{P:massiveMIMOP2} and hence that to the original problem $\PMassive$ now becomes finding the optimal number of antennas $N_1^{\star}$ to be trained, which can be straightforwardly obtained by comparing $\tilde{Q}^{\text{large-}M}_{\text{net}}(N_1)$ for $N_1\in \{0, 1, \cdots, N\}$. As evident from \eqref{eq:QnetMassiveNoTrain} and \eqref{eq:QnetMassiveWithTrain}, if $N_1\geq 1$ and $N_1\notin \mathcal{N}_{\text{large-}M}$, we always have $\tilde{Q}^{\text{large-}M}_{\text{net}}(N_1)<\tilde{Q}^{\text{large-}M}_{\text{net}}(0)$. Therefore, the optimal $N_1^{\star}$ can be obtained as
\begin{align}
N_1^{\star}=\mathrm{arg} \underset{N_1\in \{0\} \cup \mathcal{N}_{\text{large-}M} }{\max} \ \tilde{Q}_{\text{net}}^{\text{large-}M}(N_1).\label{eq:N1StarMassive}
\end{align}

Similar observations to those  in Section~\ref{sec:Rayleigh} and Section~\ref{sec:MISO} can be made based on \eqref{eq:setNMassive} and \eqref{eq:N1StarMassive}, i.e., for WET in massive MIMO Rician fading channels, channel training helps if $T$ and $\Gamma$ are sufficiently large, and $K$ is sufficiently small.

To gain further insights, we consider the case where the deterministic Rician channel component $\Hbar$  is given by a rank-1 matrix.  This corresponds to a propagation environment without any fixed scatterer cluster, so that the deterministic Rician channel component is contributed by the LOS path only. In this case, $L=1$ in \eqref{eq:Hbar} and hence $\Hbar$ can be represented as
\begin{align}\label{eq:HbarRank1}
\Hbar=\mathbf a_r(\theta_r) \mathbf a_t(\theta_t)^H.
\end{align}
\begin{lemma}\label{lemma:scalewithM}
For WET systems in MIMO Rician fading channels with rank-1 deterministic channel component and sufficiently large channel block length $T$ satisfying $T>KN^2+2N$, the net average harvested energy $\bar{Q}_{\text{net}}$ with the proposed training-based scheme scales with $M$ asymptotically as
\begin{align}\label{eq:scalewithM}
\bar{Q}_{\text{net}}\geq \tilde{Q}_{\text{net}}^{\text{large-}M} \rightarrow (T-N)\eta P_f \beta \frac{KN+1}{K+1}M, \text{ as } M\rightarrow \infty.
\end{align}
\end{lemma}

\begin{IEEEproof}
Please refer to Appendix~\ref{A:scalewithM}.
\end{IEEEproof}

As a comparison, we consider in the following the asymptotic scaling for the two benchmark schemes, i.e., the ideal energy beamforming assuming perfect CSI at the ET and the energy beamforming based on the deterministic LOS channel component only.

\begin{lemma}\label{lemma:scaleIdealLOS}
For WET systems in MIMO Rician fading channels with rank-1 deterministic channel component, the average harvested energy with ideal energy beamforming and that based on the LOS channel component behave as $M\rightarrow \infty$  as
\begin{align}
\bar{Q}^{\text{ideal}}&\rightarrow T \eta P_f \beta \frac{KN+1}{K+1}M, \label{eq:scaleIdeal}\\
\bar{Q}^{\text{LOS}}&\rightarrow T \eta P_f \beta \frac{KN}{K+1}M \label{eq:scaleLOS}.
\end{align}
\end{lemma}
\begin{IEEEproof}
The proof follows from the similar techniques used in \eqref{eq:massiveMIMOApprox}, and hence is omitted here for brevity.
\end{IEEEproof}

The following result immediately follows from Lemma~\ref{lemma:scalewithM} and Lemma~\ref{lemma:scaleIdealLOS}.
\begin{proposition}\label{cor:sameScale}
For WET systems in MIMO Rician fading channels with rank-1 deterministic channel component and sufficiently large channel block length $T\gg N$, the net average harvested energy achieved by the optimized training scheme has the same asymptotic scaling with $M$ as that achieved by ideal energy beamforming.
\end{proposition}

Note that the condition $T\gg N$ in Proposition~\ref{cor:sameScale} generally holds for practical WET systems, since $N$ is usually quite small due to the limited space available at the ER (e.g., wireless sensor nodes), whereas $T$ is usually on the order of hundreds or even thousands in slow fading environment. It is also observed from \eqref{eq:scaleIdeal} and \eqref{eq:scaleLOS} that the optimal asymptotic scaling cannot be achieved by the simple LOS-based energy beamforming, except for the extreme case with $K\rightarrow \infty$.

\begin{figure}
\centering
\includegraphics[scale=0.5]{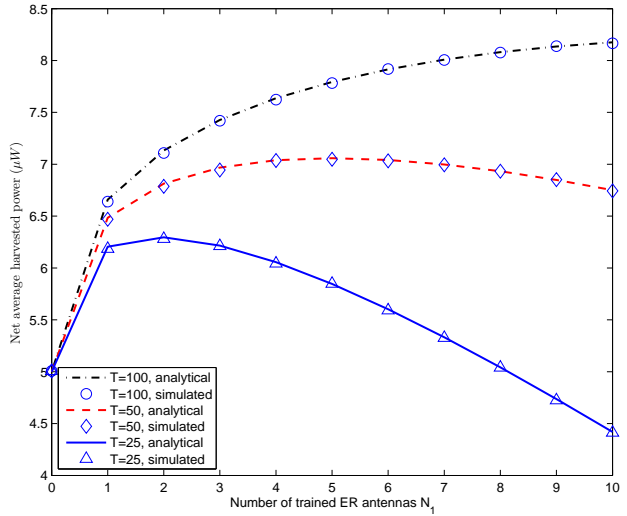}
\caption{Net average harvested power versus the number of trained ER antennas $N_1$ for different block lengths $T$, with $M=5$ and $N=10$ in MIMO Rayleigh fading channel.}\label{F:QnetVsN1}
\end{figure}

\section{Numerical Results}\label{sec:numerical}
In this section, numerical examples are provided to corroborate our study. We assume that the ET has a maximum transmit power of $P_f=1$ Watt. The average signal power attenuation  is assumed to be $60$ dB, i.e., $\beta=10^{-6}$, which may correspond to an operating distance between the ET and ER around $30$ meters with carrier frequency $900$ MHz. Furthermore, the noise power received at the ET during training phase is assumed be $\sigma_r^2=-90$ dBm and the energy harvesting efficiency at the ER is $\eta=0.5$. 
 For the channel model given in \eqref{eq:HRician}, the deterministic component $\Hbar$ is assumed to be rank-1 as given by \eqref{eq:HbarRank1}, with AOA and AOD respectively given by $\theta_r=0^{\circ}$ and $\theta_t=10^\circ$. Furthermore, for simplicity, we assume that ULAs with adjacent elements separated by half wavelength are deployed at the ER and ET, so that the phase differences among  the antenna elements are given in \eqref{eq:Phim}.
 In the following, we present the numerical results for the three cases of MIMO Rayleigh fading channel, MISO and massive MIMO Rician fading channels, respectively.

\subsection{MIMO Rayleigh Fading Channel}
First, we consider a WET system in MIMO Rayleigh fading channel with $M=5$ and $N=10$. In Fig.~\ref{F:QnetVsN1}, by varying the number of trained ER antennas $N_1$, the net average harvested energy  normalized by the channel block length $T$ is plotted for different $T$ values, where the average is taken over $10000$ random channel realizations. The analytical results derived in Section~\ref{sec:Rayleigh}, i.e., $\bar{Q}_{\text{net}}^{\mathrm{R}}(N_1)/T$ with $\bar{Q}_{\text{net}}^{\mathrm{R}}(N_1)$ given by \eqref{eq:QnetN1Equal0} for $N_1=0$ or by \eqref{eq:QnetN1Greater0} for $1\leq N_1\leq N$,  are also shown in the same figure. It is observed that the simulation and analytical results have perfect match with each other, which thus validates our theoretical studies. Furthermore, for moderate block lengths of $T=25$ and $T=50$, Fig.~\ref{F:QnetVsN1} clearly shows the trade-offs in selecting the number of ER antennas to be trained. It is observed that the optimal number of training ER antennas is  $2$ for $T=25$ and increases to $5$ for $T=50$. As the block length $T$ increases to $100$ symbol durations, the net average harvested power  monotonically increases with $N_1$, and hence it is optimal to train all of the $N=10$ available ER antennas.

\begin{figure}
\centering
\includegraphics[scale=0.5]{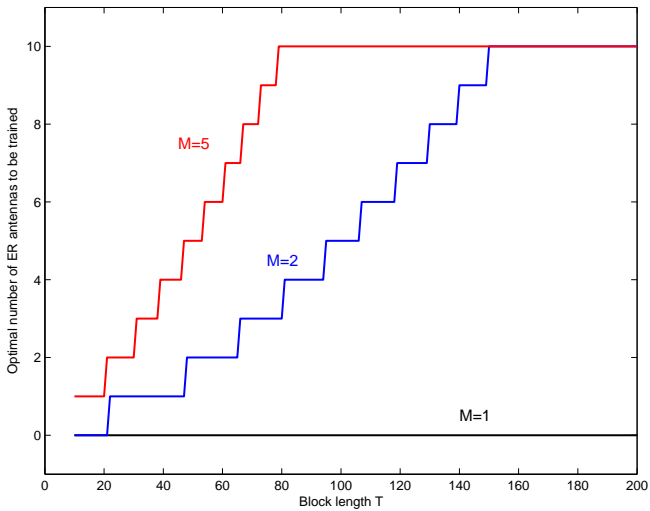}
\caption{Optimal number of trained  ER antennas versus block length $T$ for different number of ET antennas $M$ with $N=10$ in MIMO Rayleigh fading channels.\vspace{-3ex}}\label{F:N1StarVsT}
\end{figure}

\begin{figure}
\centering
\includegraphics[scale=0.5]{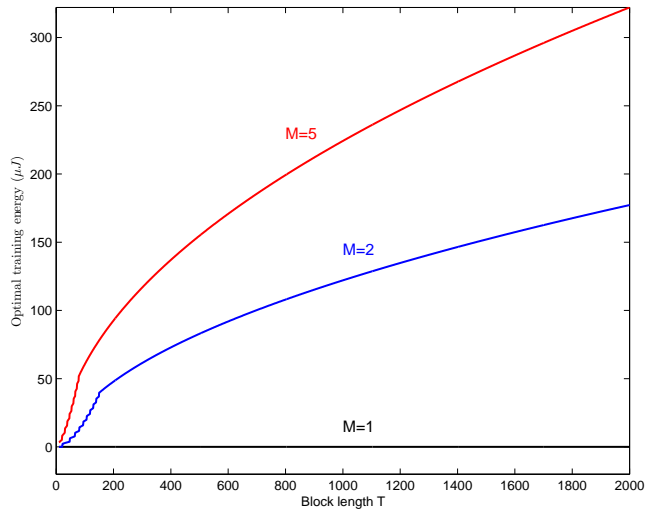}
\caption{Optimal training energy versus block length $T$ for different number of ET antennas $M$ with $N=10$ in MIMO Rayleigh fading channels.\vspace{-3ex}}\label{F:ErStarVsT}
\end{figure}

  In Fig.~\ref{F:N1StarVsT}, the optimal number of ER antennas $N_1^{\star}$ that should be trained is plotted against the channel block length $T$ for different number of ET antennas $M$. The total number of antennas available at the ER is $N=10$. Note that based on Lemma~\ref{lemma:tauEqualN1}, $N_1^{\star}$ also gives the optimal number of symbol durations $\tau^{\star}$ that should be allocated for training during each coherent block. It is observed that for $M=1$, no training should be applied regardless of the block length $T$. This is expected since no beamforming gain can be exploited when there is only one antenna at the ET. In contrast, when the ET has multiple antennas, e.g., $M=2$ or $M=5$, the optimal number of trained  ER antennas increases with the block length $T$.  As $T$ becomes sufficiently large, i.e., $T\geq 150$ for $M=2$ and $T\geq 80$ for $M=5$, all antennas at the ER should be trained. It is also observed from Fig.~\ref{F:N1StarVsT} that more ER antennas should be trained for $M=5$ than that for $M=2$, which is expected since transmit beamforming is more effective and hence training is more beneficial when more antennas are available at the ET. 
 Similar observations can be made in Fig.~\ref{F:ErStarVsT}, where the optimal training energy $E_r^{\star}$ versus the channel block length $T$ for the same setups as in Fig.~\ref{F:N1StarVsT} is shown. 
 It is noted that for sufficiently large $T$, $E_r^{\star}$ increases with $T$ in a square-root relationship, which implies a diminishing marginal gain  offered by channel training as $T$ increases.

\begin{figure}
\centering
\includegraphics[scale=0.5]{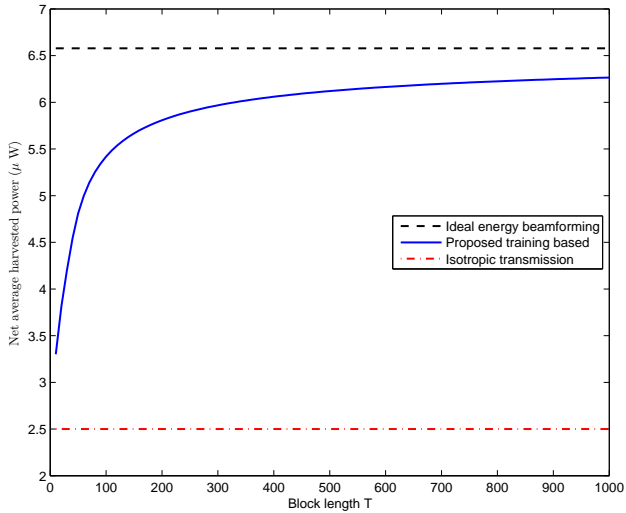}
\caption{Net average harvested power versus block length $T$ for WET in MIMO Rayleigh fading channel with $M=N=5$.\vspace{-3ex}}\label{F:QnetVsT}
\end{figure}
In Fig.~\ref{F:QnetVsT}, the net average harvested power of the optimized training-based scheme is plotted against the block length $T$ in a MIMO Rayleigh fading channel with $M=N=5$. The performances of two benchmark schemes, namely the ideal energy beamforming assuming perfect CSI at the ET and the isotropic transmission (i.e., $\mathbf S=(P_f/M)\mathbf 1 \mathbf 1^H$ with $\mathbf 1$ denoting an all-one vector) with no CSI, are also plotted. It is observed that the training-based WET significantly outperforms isotropic transmission, and its performance improves with the increasing of $T$. This is expected since for larger  block length, it is affordable to have more training as shown in Fig.~\ref{F:N1StarVsT} and Fig.~\ref{F:ErStarVsT}, and hence the channel can be more accurately estimated at the ET and the obtained  energy beamforming is more effective. It is also observed that with sufficiently large  $T$, the training-based WET approaches the performance upper bound with perfect CSI.

\begin{figure}
\centering
\includegraphics[scale=0.5]{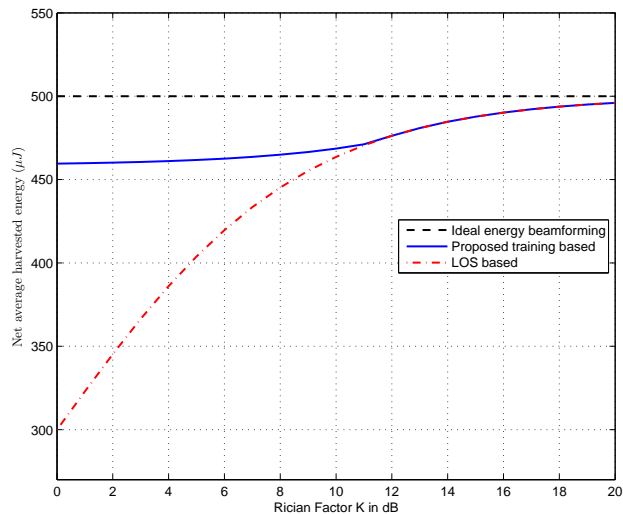}
\caption{Net average harvested energy versus the Rician factor $K$ for WET  in MISO Rician channel with $M=5$ and $T=200$.\vspace{-3ex}}\label{F:QnetVsK}
\end{figure}

\subsection{MISO Rician Fading Channel}
Next, we consider the WET systems in MISO Rician fading channels with $N=1$ and $K>0$. In Fig.~\ref{F:QnetVsK}, the net average harvested energy with the proposed training-based scheme is plotted against the Rician factor $K$. The number of ET antennas and the channel block length are respectively set as $M=5$ and $T=200$. The  two benchmark schemes are also shown in the same figure, which are the ideal energy beamforming assuming perfect CSI at the ET and that solely based on the LOS component of the channel (i.e., $\mathbf S=P_f \bar{\mathbf v} \bar{\mathbf v}^H$, with $\bar{\mathbf v}$ denoting the dominant eigenvector of $\Hbar^H\Hbar$). It is observed that the proposed training-based WET significantly outperforms the LOS-based energy beamforming scheme for small $K$ values, for which the power of the scattered signal component is comparable to that of the LOS component. In such scenarios, channel training greatly improves the accuracy of the CSI at the ET, which thus makes energy beamforming more effective than that based on the LOS component only. As $K$ increases, and hence the LOS component becomes dominating, the performance gap between the two schemes diminishes. With sufficiently large $K$, it is observed that the proposed scheme degenerates to the LOS-based energy beamforming, which approaches to the ideal scenario with perfect CSI at the ET.

\begin{figure}
\centering
\includegraphics[scale=0.5]{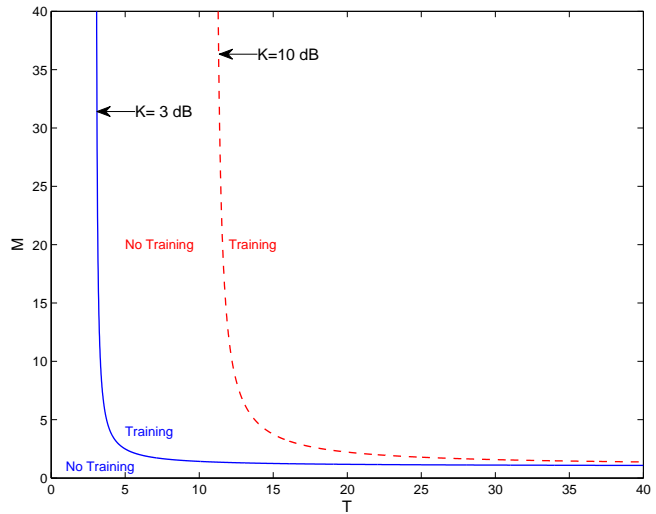}
\caption{Training and non-training regions for WET in MISO Rician fading channels on the $M$-versus-$T$ plane.\vspace{-3ex}}\label{F:trainingRegionRician}
\end{figure}

In Fig.~\ref{F:trainingRegionRician}, the optimal training and no-training regions  based on \eqref{eq:N1OptMISO} are plotted in the $M$-versus-$T$ plane for $K=3$ dB and $K=10$ dB with sufficiently large $\Gamma$. Fig.~\ref{F:trainingRegionRician} affirms our conclusion drawn in Section~\ref{sec:MISO} that for fixed Rician factor $K$, training helps if and only if the channel block length $T$ and the number of ET antennas $M$ are sufficiently large. It is also observed from Fig.~\ref{F:trainingRegionRician} that as the Rician factor $K$ decreases, the training region enlarges, which is as expected.

\subsection{Massive MIMO Rician Fading Channel}
\begin{figure}
\centering
\includegraphics[scale=0.5]{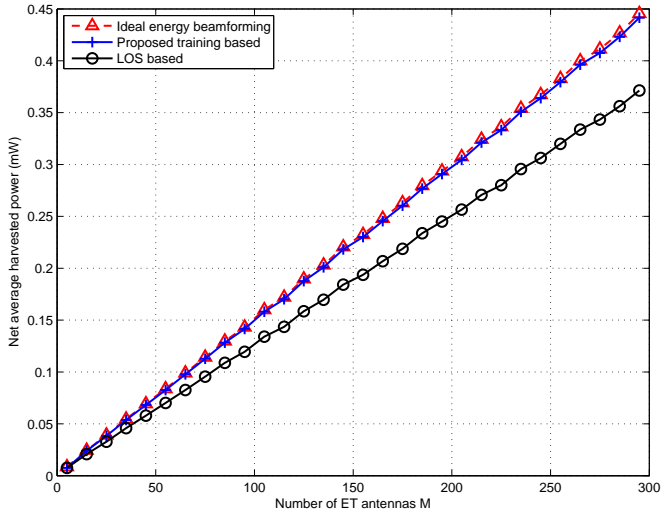}
\caption{Net average harvested power versus number of ET antennas $M$ for WET in MIMO Rician fading channels with $K=1$, $T=1000$ and $N=5$.\vspace{-3ex}}\label{F:QnetVsMMassiveMIMO}
\end{figure}

Last, we consider the effect of large $M$ on the WET systems in massive MIMO Rician channels as discussed in Section~\ref{sec:massive}.  In Fig.~\ref{F:QnetVsMMassiveMIMO}, the net average harvested energy normalized by channel block length $T$ is plotted against the number of ET antennas $M$, with $M$ ranging from $5$ to $300$.  The channel block length and Rician factor are respectively set as $T=1000$ and $K=1$. Furthermore, the number of antennas at the ER is set as $N=5$. The results for the two benchmark schemes with ideal energy beamforming and LOS-based beamforming are also shown in the figure.  It is observed from Fig.~\ref{F:QnetVsMMassiveMIMO} that, the proposed training design based on massive MIMO setup achieves a very close performance as the ideal energy beamforming, even for moderate $M$ values.  In contrast, the simple LOS-based energy beamforming performs significantly worse than the other two schemes, and the performance gap increases as $M$ gets larger. This implies the necessity for channel training in MIMO WET systems to exploit the scattered channel power, especially when $M$ is large and the potential beamforming gain is significant.

\section{Conclusion and Future Work}\label{sec:conclusion}
This  paper studied the optimal design of channel training for MIMO WET systems in Rician fading channels. Assuming channel reciprocity, the forward link channel of WET can be efficiently estimated at the ET based on the training signals sent by the ER in the reverse communication link. To optimally resolve the tradeoff between channel training and the resulting energy beamforming gain, we proposed a novel design framework to maximize the \emph{net} harvested energy at the ER to explicitly take into account the time and energy costs for channel training. A training design optimization problem was formulated for the general uncorrelated MIMO Rician fading channels, while the optimal solutions were derived for the special cases of MIMO Rayleigh and MISO Rician fading channels. For massive MIMO Rician fading channels, a high-quality approximate solution was obtained, which is shown to achieve  the optimal asymptotic scaling with the increasing number of ET antennas.

There are a number of research directions along which the developed results in this paper  can be further investigated, as briefly discussed as follows.

\begin{itemize}
\item \emph{Correlated Fading Channel:} It is interesting to investigate the optimal training design for correlated MIMO Rician fading channels, which provide more accurate modeling in poor scattering environment and/or when the antennas are not sufficiently separated. In this case, the structure of the training sequence needs to be optimized as well, since orthogonal training sequence may not be optimal in general for correlated MIMO fading channels \cite{533}.
\item \emph{Multi-User Setup:} More  research endeavor is needed to find the optimal training design for the general multi-user setups with near-far ERs. In particular, as evident from \eqref{eq:Gamma}, the optimal solution needs to resolve the so-called ``doubly near-far'' problem, where a far ER from the ET suffers from higher propagation loss than a near  ER for both reverse-link channel training and forward-link energy transmission. Note that this phenomenon has been first revealed in \cite{515} for WPCN.
\item \emph{Energy Outage:} In this paper, the net average harvested energy at the ER is maximized by optimizing the training design. In certain applications, a more appropriate performance metric may be the energy outage probability, the minimization of which deserves further  study.
\end{itemize}


\appendices
\section{Proof of Lemma~\ref{lemma:tauEqualN1}}\label{A:tauEqualN1}
 Lemma~\ref{lemma:tauEqualN1} can be shown by contradiction. Suppose, on the contrary, that the optimal solution  $(N_1^\star, \tau^\star, P_r^\star)$ to $\PRayleigh$ satisfies $\tau^{\star}>N_1^{\star}$. Then we define a new tuple $(N_1', \tau', P_r')$, with $N_1'=N_1^\star$, $\tau'=N_1^\star$, and $P_r'=P_r^{\star}\tau^{\star}/N_1^{\star}$. In other words, in the newly defined training design, the number of training antennas $N_1$ and the total training energy $P_r\tau$ keep unchanged, while the training duration is decreased to save the time overhead. It can be verified that with such a construction,  $(N_1', \tau', P_r')$ is feasible to $\PRayleigh$. Furthermore, due to the reduced time overhead, it follows from \eqref{eq:QnetRayleigh} that the following inequality holds: 
\begin{align}
\bar Q^{\mathrm{R}}_{\text{net}}\left(N_1', \tau', P_r'\right) > \bar Q^{\mathrm{R}}_{\text{net}}(N_1^{\star},  \tau^{\star}, P_r^{\star}).\label{eq:strcitImprv}
\end{align}

In other words, there exists another training design $(N_1', \tau', P_r')$ as defined above, which satisfies  the constraints of $\PRayleigh$ 
and strictly increases the objective value $\bar Q^{\mathrm{R}}_{\text{net}}$. This thus contradicts the assumption that $(N_1^{\star}, \tau^{\star}, P_r^{\star})$  is  optimal. Therefore, the assumption $\tau^\star >N_1^\star$ is invalid, or  $\tau^{\star}\leq  N_1^{\star}$ must hold. Together with the constraint $\tau\geq N_1$ in $\PRayleigh$, we thus have $\tau^{\star}=N_1^{\star}$ for the optimal solution to $\PRayleigh$.

This completes the proof of Lemma~\ref{lemma:tauEqualN1}.

\section{Proof of Lemma~\ref{lemma:scalewithM}}\label{A:scalewithM}
To show Lemma~\ref{lemma:scalewithM}, we need to first obtain the optimal number of ER antennas $N_1^{\star}$ to be trained by solving \eqref{eq:N1StarMassive}.
 With rank-1 deterministic channel component $\Hbar$ given in \eqref{eq:HbarRank1}, the solution to \eqref{eq:N1StarMassive} can be obtained in closed-form, as we show next. With \eqref{eq:HbarRank1}, it can be shown  that the dominant eigenvalue $\bar{\lambda}$ and the corresponding eigenvector $\bar{\mathbf v}$ of $\Hbar^H\Hbar$ are respectively given by
 \begin{align}
\bar{\lambda}&=\lambda_{\max}(\Hbar\Hbar^H)=MN, \label{eq:lambdaBar}\\
\bar{\mathbf v}&=\mathbf v_{\max}(\Hbar\Hbar^H)=\frac{1}{\sqrt{N}} \mathbf a_r(\theta_r).
 \end{align}
 It follows from \eqref{eq:ar} that for any AOA $\theta_r$, all elements in the vector $\mathbf a_r(\theta_r)$ have absolute values equal to $1$. As a result, we have
 \begin{align}
 \|[\bar{\mathbf v}^{\star}]_{1:N_1}\|^2=\frac{N_1}{N}, \ 1\leq N_1\leq N.\label{eq:vstarNorm}
 \end{align}
 Furthermore, in the massive MIMO regime with $M\gg N$ and with moderate values for $K$ and $\Gamma$, we have
 \begin{align}
 \sqrt{(T-N_1)\left(\frac{M}{N}-1\right)} \gg \frac{K+1}{\sqrt{\Gamma}}, \ \forall 1\leq N_1\leq N. \label{eq:muchLarger}
 \end{align}
 With \eqref{eq:lambdaBar}, \eqref{eq:vstarNorm}, and \eqref{eq:muchLarger}, by keeping only the dominant terms,  the expressions of $\tilde{Q}^{\text{large-}M}_{\text{net}}(N_1)$ for $N_1=0$ given in \eqref{eq:QnetMassiveNoTrain} and for $1\leq N_1\leq N$ given in \eqref{eq:QnetMassiveWithTrain} can be unified as
 \begin{align}\label{eq:QnetLargeMApprox}
 \hspace{-3ex} \tilde{Q}^{\text{large-}M}_{\text{net}}(N_1) \approx
\frac{\eta P_f \beta}{K+1}(T-N_1) \left(KN + \frac{N_1}{N}\right)M,  \ 0\leq N_1 \leq N.
 \end{align}

 As a consequence, problem \eqref{eq:N1StarMassive} reduces to
 \begin{align}\label{eq:N1starMassiveReduced}
 N_1^{\star}=\arg \underset{0\leq N_1 \leq N, N_1\in \mathbb{Z}}{\max}\ \frac{\eta P_f \beta}{K+1}(T-N_1) \left(KN + \frac{N_1}{N}\right)M.
 \end{align}

 To solve \eqref{eq:N1starMassiveReduced}, we first obtain the solution $\hat{N}_1$ to the relaxed problem without considering the integer constraint $N_1 \in \mathbb{Z}$. Then $N_1^{\star}$ can be obtained as
 \begin{equation*}
 \small
 \begin{aligned}
  N_1^{\star}=\begin{cases}
 \hat{N}_1, & \text{ if } \hat{N}_1\in \mathbb{Z},\\
 \arg \max \left\{ \tilde{Q}^{\text{large-}M}_{\text{net}}\big(\lfloor \hat N_1 \rfloor \big), \tilde{Q}^{\text{large-}M}_{\text{net}}\big(\lceil \hat N_1 \rceil \big)\right\}, & \text{ otherwise}.
 \end{cases}
 \end{aligned}
 \end{equation*}

 To capture the most essential insight, we assume that $\hat{N}_1\in \mathbb{Z}$ holds. Then the optimal solution $N_1^{\star}$ to \eqref{eq:N1starMassiveReduced} can be easily obtained as
 \begin{align}
 N_1^{\star}=\left[ \frac{T-KN^2}{2}\right]_{0}^N, \label{eq:N1StarMassiveSimplified}\emph{}
 \end{align}
 where $\left[x\right]_a^b\triangleq \max\{\min\{x,b\},a\}$.
In particular, for sufficiently large channel block length $T$ satisfying $T>KN^2+2N$, we have $N_1^{\star}=N$, i.e., all ER antennas should be trained. By substituting $N_1=N$ into \eqref{eq:QnetLargeMApprox}, the result \eqref{eq:scalewithM} in Lemma~\ref{lemma:scalewithM} follows.

This thus completes the proof of Lemma~\ref{lemma:scalewithM}.
\bibliographystyle{IEEEtran}
\bibliography{IEEEabrv,IEEEfull}

\begin{thebibliography}{10}
\providecommand{\url}[1]{#1}
\csname url@samestyle\endcsname
\providecommand{\newblock}{\relax}
\providecommand{\bibinfo}[2]{#2}
\providecommand{\BIBentrySTDinterwordspacing}{\spaceskip=0pt\relax}
\providecommand{\BIBentryALTinterwordstretchfactor}{4}
\providecommand{\BIBentryALTinterwordspacing}{\spaceskip=\fontdimen2\font plus
\BIBentryALTinterwordstretchfactor\fontdimen3\font minus
  \fontdimen4\font\relax}
\providecommand{\BIBforeignlanguage}[2]{{%
\expandafter\ifx\csname l@#1\endcsname\relax
\typeout{** WARNING: IEEEtran.bst: No hyphenation pattern has been}%
\typeout{** loaded for the language `#1'. Using the pattern for}%
\typeout{** the default language instead.}%
\else
\language=\csname l@#1\endcsname
\fi
#2}}
\providecommand{\BIBdecl}{\relax}
\BIBdecl

\bibitem{502}
H.~J. Visser and R.~J.~M. Vullers, ``{RF} energy harvesting and transport for
  wireless sensor network applications: Principles and requirements,''
  \emph{Proceedings of the IEEE}, vol. 101, no.~6, pp. 1410--1423, Jun. 2013.

\bibitem{525}
S.~Bi, C.~K. Ho, and R.~Zhang, ``Wireless powered communication: opportunities
  and challenges,'' to appear in {\it IEEE Commun. Mag.}, available online at
  http://arxiv.org/abs/1408.2335.

\bibitem{534}
X.~Lu, P.~Wang, D.~Niyato, D.~I. Kim, and Z.~Han, ``{Wireless networks with RF
  energy harvesting: a contemporary survey},'' to appear in \emph{IEEE Commun.
  Surveys Tuts.}, available online at http://arxiv.org/abs/1406.6470.

\bibitem{478}
R.~Zhang and C.-K. Ho, ``{MIMO} broadcasting for simultaneous wireless
  information and power transfer,'' \emph{{IEEE} Trans. Wireless Commun.},
  vol.~12, no.~5, pp. 1989--2001, May 2013.

\bibitem{505}
W.~C. Brown, ``The history of power transmission by radio waves,'' \emph{IEEE
  Trans. on Microwave Theory and Techniques}, vol. MTT-32, no.~9, pp.
  1230--1242, Sep. 1984.

\bibitem{506}
------, ``Experiments involving a microwave beam to power and position a
  helicopter,'' \emph{IEEE Trans. on Aerospace and Electronic Sys.}, vol.
  AES-5, no. 692-702, pp. 692--702, Sep. 1969.

\bibitem{507}
J.~O. Mcspadden and J.~C. Mankins, ``Space solar power programs and microwave
  wireless power transmission technology,'' \emph{IEEE Microw. Mag.}, vol.~3,
  no.~4, pp. 46--57, Dec. 2002.

\bibitem{508}
K.~Huang and V.~K.~N. Lau, ``Enabling wireless power transfer in cellular
  networks: Architecture, modeling and deployment,'' \emph{{IEEE} Trans.
  Wireless Commun.}, vol.~13, no.~2, pp. 902--912, Feb. 2014.

\bibitem{510}
S.~Lee, R.~Zhang, and K.~Huang, ``Opportunistic wireless energy harvesting in
  cognitive radio networks,'' \emph{{IEEE} Trans. Wireless Commun.}, vol.~12,
  no.~9, pp. 4788--4799, Sep. 2013.

\bibitem{509}
X.~Lu, P.~Wang, D.~Niyato, and E.~Hossain, ``Dynamic spectrum access in
  cognitive radio networks with {RF} energy harvesting,'' \emph{IEEE Wireless
  Commun.}, pp. 102--110, Jun. 2014.

\bibitem{535}
A.~A. Nasir, X.~Zhou, S.~Durrani, and R.~A. Kennedy, ``Relaying protocols for
  wireless energy harvesting and information processing,'' \emph{{IEEE} Trans.
  Wireless Commun.}, vol.~12, no.~7, pp. 3622--3636, Jul. 2013.

\bibitem{511}
Z.~Ding, S.~M. Perlaza, I.~Esnaola, and H.~V. Poor, ``Power allocation
  strategies in energy harvesting wireless cooperative networks,'' \emph{{IEEE}
  Trans. Wireless Commun.}, vol.~13, no.~2, pp. 846--860, Feb. 2014.

\bibitem{514}
X.~Zhou, R.~Zhang, and C.~K. Ho, ``Wireless information and power transfer:
  architecture design and rate-energy tradeoff,'' \emph{{IEEE} Trans. Commun.},
  vol.~61, no.~11, pp. 4757--4767, Nov. 2013.

\bibitem{521}
L.~Liu, R.~Zhang, and K.~C. Chua, ``Wireless information transfer with
  opportunistic energy harvesting,'' \emph{{IEEE} Trans. Wireless Commun.},
  vol.~12, no.~1, pp. 288--300, Jan. 2013.

\bibitem{522}
D.~W.~K. Ng, E.~S. Lo, and R.~Schober, ``Wireless information and power
  transfer: Energy efficiency optimization in {OFDMA} systems,'' \emph{{IEEE}
  Trans. Wireless Commun.}, vol.~12, no.~12, pp. 6352--6370, Dec. 2013.

\bibitem{515}
H.~Ju and R.~Zhang, ``Throughput maximization in wireless powered communication
  networks,'' \emph{{IEEE} Trans. Wireless Commun.}, vol.~13, no.~1, pp.
  418--428, Jan. 2014.

\bibitem{516}
L.~Liu, R.~Zhang, and K.~C. Chua, ``Multi-antenna wireless powered
  communication with energy beamforming,'' to appear in \emph{IEEE Trans.
  Commun.}, available online at http://arxiv.org/abs/1312.1450.

\bibitem{518}
G.~Yang, C.~K. Ho, R.~Zhang, and Y.-L. Guan, ``Throughput optimization for
  massive {MIMO} systems powered by wireless energy transfer,'' submitted to
  \emph{IEEE J. Sel. Areas Commun.}, available online at
  http://arxiv.org/abs/1403.3991.

\bibitem{517}
D.~Snoonian, ``Smart buildings,'' \emph{IEEE Spectrum}, vol.~40, no.~8, pp.
  18--23, Aug. 2003.

\bibitem{374}
F.~Rusek, D.~Persson, B.~K. Lau, E.~G. Larsson, T.~L. Marzetta, O.~Edfors, and
  F.~Tufvesson, ``Scaling up {MIMO}: Opportunities and challenges with very
  large arrays,'' \emph{{IEEE} Signal Process. Mag.}, vol.~30, no.~1, pp.
  40--60, Jan. 2013.

\bibitem{497}
L.~Lu, G.~Y. Li, A.~L. Swindlehurst, A.~Ashikhin, and R.~Zhang, ``An overview
  of massive {MIMO}: benefits and challenges,'' \emph{IEEE J. Sel. Topics
  Signal Process.}, vol.~8, no.~5, pp. 742--758, Oct. 2014.

\bibitem{495}
G.~Yang, C.~K. Ho, and Y.-L. Guan, ``Dynamic resource allocation for
  multiple-antenna wireless power transfer,'' \emph{{IEEE} Trans. Signal
  Process.}, vol.~62, no.~14, pp. 3565 -- 3577, Jun. 2014.

\bibitem{500}
X.~Chen, C.~Yuen, and Z.~Zhang, ``Wireless energy and information transfer
  tradeoff for limited-feedback multiantenna systems with energy beamforming,''
  \emph{{IEEE} Trans. Veh. Technol.}, vol.~63, no.~1, pp. 407--412, Jan. 2014.

\bibitem{547}
X.~Zhou, ``Training-based {SWIPT}: optimal power splitting at the receiver,''
  to appear in \emph{IEEE Trans. Veh. Technol.}, available online at
  http://arxiv.org/abs/1405.4623.

\bibitem{484}
D.~J. Love, R.~W. Heath~Jr., V.~K.~N. Lau, D.~Gesbert, B.~D. Rao, and
  M.~Andrews, ``An overview of limited feedback in wireless communication
  systems,'' \emph{{IEEE} J. Sel. Areas Commun.}, vol.~26, no.~8, pp.
  1341--1365, Oct. 2008.

\bibitem{491}
J.~Xu and R.~Zhang, ``Energy beamforming with one-bit feedback,'' \emph{{IEEE}
  Trans. Signal Process.}, vol.~62, no.~20, pp. 5370--5381, Oct. 2014.

\bibitem{373}
T.~L. Marzetta, ``Noncooperative cellular wireless with unlimited numbers of
  base station antennas,'' \emph{{IEEE} Trans. Wireless Commun.}, vol.~9,
  no.~11, pp. 3590--3600, Nov. 2010.

\bibitem{499}
F.~R. Farrokhi, G.~J. Foschini, A.~Lozano, and R.~A. Valenzuela, ``Link-optimal
  space-time processing with multiple transmit and receive antennas,''
  \emph{{IEEE} Commun. Lett.}, vol.~5, no.~3, pp. 85--87, Mar. 2001.

\bibitem{519}
H.~Bolcskei, M.~Borgmann, and A.~J. Paulraj, ``Impact of the propagation
  environment on the performance of space-frequency coded {MIMO-OFDM},''
  \emph{{IEEE} J. Sel. Areas Commun.}, vol.~21, no.~3, pp. 427--439, Apr. 2003.

\bibitem{440}
A.~Forenza, D.~J. Love, and R.~W. Heath~Jr., ``Simplified spatial correlation
  models for clustered {MIMO} channels with different array configurations,''
  \emph{{IEEE} Trans. Veh. Technol.}, vol.~56, no.~4, pp. 1924--1934, Jul.
  2007.

\bibitem{498}
L.~G. Ordonez, D.~P. Palomar, and J.~R. Fonollosa, ``Ordered eigenvalues of a
  general class of {H}ermitian random matrices with application to the
  performance analysis of {MIMO} systems,'' \emph{{IEEE} Trans. Signal
  Process.}, vol.~57, no.~2, pp. 672--689, Feb. 2009.

\bibitem{2}
B.~Hassibi and B.~M. Hochwald, ``How much training is needed in
  multiple-antenna wireless links?'' \emph{{IEEE} Trans. Inf. Theory}, vol.~49,
  no.~4, pp. 951--963, Apr. 2003.

\bibitem{71}
S.~M. Kay, \emph{Fundumentals of Statistical Signal Processing: Estimation
  Theory.}\hskip 1em plus 0.5em minus 0.4em\relax New Jersey: Prentice-Hall,
  1993.

\bibitem{477}
P.~A. Dighe, R.~K. Mallik, and S.~S. Jamuar, ``Analysis of transmit-receive
  diversity in {R}ayleigh fading,'' \emph{{IEEE} Trans. Commun.}, vol.~51,
  no.~4, pp. 694--703, Apr. 2003.

\bibitem{476}
A.~Maaref and S.~Aissa, ``Closed-form expressions for the outage and ergodic
  {S}hannon capacity of {MIMO MRC} systems,'' \emph{{IEEE} Trans. Commun.},
  vol.~53, no.~7, pp. 1092--1095, Jul. 2005.

\bibitem{533}
J.~H. Kotecha and A.~M. Sayeed, ``Transmit signal design for optimal estimation
  of correlated {MIMO} channels,'' \emph{{IEEE} Trans. Signal Process.},
  vol.~52, no.~2, pp. 546--557, Feb. 2014.

\end{thebibliography}

\end{document}